\documentclass[%
superscriptaddress,
nofootinbib,
 amsmath,amssymb,
 aps,prd
]{revtex4-2}

\usepackage{graphicx,subfigure,ulem}
\usepackage{dcolumn}
\usepackage{bm}
\usepackage{natbib} 
\usepackage[colorlinks=true,citecolor=blue,linkcolor=blue,urlcolor=blue]{hyperref}
\setcitestyle{square, numbers, comma, sort&compress} 

\begin{document}

\title{Hot perturbative QCD in a very strong magnetic background}


\author{Eduardo S. Fraga}
\email{fraga@if.ufrj.br}
\affiliation{Instituto de F\'isica, Universidade Federal do Rio de Janeiro, Caixa Postal 68528, 21941-972, Rio de Janeiro, RJ, Brazil}

\author{Let\'icia F. Palhares}%
 \email{leticia.palhares@uerj.br}
 \affiliation{Universidade do Estado do Rio de Janeiro, Instituto de F\'isica, Departamento de F\'isica Te\'orica, Rua S\~ao Francisco Xavier 524, 20550-013 Maracan\~a, Rio de Janeiro, Brasil}
 
\author{Tulio E. Restrepo}
 \email{tulioeduardo@pos.if.ufrj.br}

\affiliation{Instituto de F\'isica, Universidade Federal do Rio de Janeiro, Caixa Postal 68528, 21941-972, Rio de Janeiro, RJ, Brazil}%


\begin{abstract}
We compute the pressure, chiral condensate and strange quark number susceptibility from first principles within perturbative QCD at finite temperature and very high magnetic fields up to two-loop and physical quark masses. The region of validity for our framework is given by $m_s \ll T \ll \sqrt{eB}$, where $m_s$ is the strange quark mass, $e$ is the fundamental electric charge, $T$ is the temperature, and $B$ is the magnetic field strength. We study the convergence of the perturbative series for the pressure for different choices of renormalization scale in the running coupling, $\alpha_s (T,B)$. Our results for the chiral condensate and strange quark number susceptibility can be directly compared to recent lattice QCD data away from the chiral transition. Even though current lattice results do not
overlap with the region of validity above, perturbative results seem to be in the same ballpark.
\end{abstract}

\maketitle

\section{Introduction}

The understanding of the phase structure of hadronic matter under the influence of different control parameters, such as temperature, baryon chemical potential and electromagnetic fields, must ultimately be derived from in-medium quantum chromodynamics (QCD), its fundamental theory. The case of magnetic QCD, where one of the control parameters is an external magnetic field, is phenomenologically relevant in different scenarios. In the astrophysics of compact stars, magnetars can exhibit very large fields, of the order of $10^{15}$ Gauss \cite{Duncan:1992hi,Thompson:1993hn,Kouveliotou:1998ze}, which corresponds to $\sim 20$ MeV$^2$. In non-central, high-energy heavy ion collisions one can reach much larger values, $\sim 10^{19}$ Gauss $\sim 10 m_\pi^2$ \cite{Kharzeev:2007jp,Skokov:2009qp,Voronyuk:2011jd,Bzdak:2011yy,Deng:2012pc,Inghirami:2016iru,Roy:2017yvg}. In the early universe, primordial magnetic fields could be a few orders of magnitude higher \cite{Vachaspati:1991nm,Enqvist:1993np,Grasso:2000wj}.

From the theoretical perspective, the case of thermal magnetic QCD, where the control parameters are the temperature $T$ and external magnetic field $B$, is particularly attractive. Since it does not suffer from the Sign Problem \cite{Aarts:2015tyj}, it can be tackled by Monte Carlo simulations, and lattice QCD has produced a variety of relevant results in the last decade, including a great portion of the phase diagram \cite{DElia:2010abb,Bali:2011qj,Ilgenfritz:2012fw,Bali:2012zg,Bornyakov:2013eya,Bali:2013esa,Bruckmann:2013oba,Bali:2014kia,Endrodi:2015oba,DElia:2018xwo}. It can also be addressed analytically within limits of the fundamental theory: in perturbation theory \cite{Blaizot:2012sd,Ayala:2014uua,Ayala:2015bgv}, for large values of $T$ and $B$; in hard thermal loop perturbation theory \cite{Rath:2017fdv,Haque:2017nxq,Karmakar:2019tdp,Bandyopadhyay:2017cle,Karmakar:2020mnj}; for a large number of colors $N_c$ \cite{Fraga:2012ev}; in the low-energy sector, via chiral perturbation theory \cite{Colucci:2013zoa,Hofmann:2020dvz,Hofmann:2020ism}. Of course, hot hadronic matter in the presence of external magnetic fields can also be described within effective models. For a detailed discussion and list of references, see Refs. \cite{Fraga:2012rr,Kharzeev:2013jha,Andersen:2014xxa,Miransky:2015ava}.

In this paper we investigate the behavior of the pressure, chiral condensate and strange quark number susceptibility from first principles  within perturbative QCD at finite temperature and very high magnetic fields up to two-loop (2L) for $3$ flavors with physical quark masses. For the pressure we show that the exchange contribution increases with the magnetic field, but nevertheless corresponds to a correction of less than 20$\%$ at intermediate temperatures ($T\sim 300$ MeV) even for extremely large magnetic fields.

In order to compare our perturbative results to the benchmark provided by lattice QCD simulations, we need very large magnetic fields on the lattice, so that the domain of validity of our calculation, given by $m_s \ll T \ll \sqrt{eB}$, where $m_s$ is the strange quark mass and $e$ is the fundamental electric charge, can be reached. A few years ago, Endr\" odi \cite{Endrodi:2015oba}, in a pioneering tour de force, was able to reach magnetic fields of the order of $eB=3.25$ GeV$^2$ in his simulations. The expectation, then, using extrapolations of the available lattice data combined with an effective description of QCD, was that for magnetic fields $eB \sim 10$ GeV$^2$ the crossover in the temperature-magnetic field phase diagram would become a true first-order phase transition.  Recently D'Elia et al. \cite{DElia:2021tfb,DElia:2021yvk} have extended thermal magnetic QCD on the lattice to magnetic fields as large as $eB=9$ GeV$^2$, providing numerical evidence that the onset of a first-order line happens within the range $eB=4-9$ GeV$^2$.

This work is organized as follows. In Section \ref{sec:pressure} we present the perturbative setup and a few details on the calculation of the pressure and chiral condensate to 2L, as well as the running of the coupling and strange quark masses. In Section \ref{sec:results} we discuss our results and compare some of them to what has been obtained recently on the lattice. Section \ref{sec:outlook} contains our summary and outlook.

\section{Pressure and chiral condensate}
\label{sec:pressure}

In this section we compute the pressure and chiral condensate in the lowest Landau level approximation up to 2L in perturbative QCD. We assume that the system is embedded in a uniform, {\it very} large magnetic field ${\bf B}=B \hat{\bf z}$, where the field strength $B$ is much larger than the temperature and all masses.

\subsection{One-loop contribution to the pressure}

Let us start with the one-loop (free), contribution to the pressure of thermal QCD in the presence of high magnetic fields. The one-loop (1L) contribution coming from the quark sector is given by the following renormalized expression (subtracting the pure vacuum term) \cite{Fraga:2012rr,Kharzeev:2013jha,Andersen:2014xxa,Miransky:2015ava}:

\begin{widetext}
\begin{align}
\begin{split}
 \frac{P_{\rm free}^q}{N_c}=&\sum_f\frac{(q_f B)^2}{2\pi^2}\left[\zeta^\prime(-1,x_f)-\zeta^\prime(-1,0)+\frac{1}{2}\left(x_f-x_f^2\right)\ln x_f+\frac{x_f^2}{2}\right]\\
 &+T\sum_{n,f}\frac{q_f B}{\pi}\left(1-\delta_{n,0}/2\right)\int \frac{dp_z}{2\pi}\bigg \{\ln\left(1+e^{-\beta\left[E(n,p_z)-\mu_f\right]}\right)+\ln\left(1+e^{-\beta\left[E(n,p_z)+\mu_f\right]}\right)\bigg \} \, ,
 \end{split}\label{P0}
 \end{align}
 \end{widetext}
where $E^2(n,p_z)=p_z^2+m_f^2+2q_f B n$, $x_f\equiv m_f^2/2q_f B$, $T=1/\beta$ is the temperature, $\mu$ is the quark chemical potential, $N_c$ is the number of colors, $f$ labels quark flavors, $q_f$ is the quark electric charge, and $n=0, 1, 2, \cdots$ stands for the Landau levels. In this expression, Matsubara sums have already been performed in the medium contribution.

One should notice that there is an inherent arbitrariness in the renormalization procedure (see Refs.\cite{Fraga:2008qn,Mizher:2010zb,Fraga:2012fs,Endrodi:2013cs,Haber:2014ula,Avancini:2020xqe,Tavares:2021fik,Farias:2021fci} for a discussion). In Eq. (\ref{P0}), all mass-independent terms were neglected and the pure magnetic term goes to zero in the limit $m\to 0$. There are renormalization procedures where other terms survive and the pure magnetic expression diverges as $m\to 0$. This discrepancy in the renormalized expression leads to differences in some physical quantities, e.g. the magnetization \cite{Endrodi:2013cs}. However, it turns out that the two different pure magnetic terms have the same derivative with respect to the mass, so that quantities such as the condensate and the self-energy must in principle coincide in both approaches.

Taking the limit of very high magnetic fields ($m_s \ll T \ll \sqrt{eB}$), one ends up with the lowest Landau level (LLL) expression
\begin{widetext}
\begin{align}
\begin{split}
 \frac{P_{\rm free}^{\rm LLL}}{N_c}=&
 -\sum_f\frac{(q_fB)^2}{2\pi^2}\left[x_f\ln\sqrt{x_f}\right]+T\sum_{f}\frac{q_fB}{2\pi}\int \frac{dp_z}{2\pi}\bigg \{\ln\left(1+e^{-\beta\left[E(0,p_z)-\mu_f\right]}\right)+\ln\left(1+e^{-\beta\left[E(0,p_z)+\mu_f\right]}\right)\bigg \} \, .
 \end{split} \label{Pfree}
\end{align}
\end{widetext}

The 1L contribution from the gluons has the usual Stefan-Boltzmann form \cite{Kapusta:2006pm}
\begin{equation}
P_{\rm free}^G=2(N_c^2-1)\frac{\pi^2 T^4}{90} \,.
\end{equation}

\subsection{Two-loop contribution to the pressure}

The 2L contribution from the quark sector to the pressure of thermal QCD in the presence of high magnetic fields was computed in Ref. \cite{Blaizot:2012sd}. For numerical purposes, however, it is convenient to recast the result found in that reference in a different fashion. 

Let us start with the 2L (exchange) pressure, in the LLL approximation extracted from Ref. \cite{Blaizot:2012sd}, 
 \begin{align}
\frac{P_{\rm exch}^{\rm LLL}}{N_c}=-\frac{1}{2}\left(\frac{q_fB}{2\pi}\right)\int \frac{dk_1 dk_2}{(2\pi)^2} e^{-\frac{k_1^2+k_2^2}{2 q_f B}}
  \mathcal{G}(m_k^2=k_1^2+k_2^2,m_f^2) \, ,
  \label{P_exch_ini}
 \end{align}

where
\begin{widetext}
\begin{align}
 \mathcal{G}(m_k^2,m_f^2)=& g^2 \left(\frac{N_c^2-1}{2}\right)\int\frac{dk_zdp_zdq_z}{(2\pi)^3}(2\pi)\delta(p_z-q_z-k_z)T^3\sum_{\ell,n_1,n_2}\beta\delta_{n_1,n_2+\ell}\frac{4m_f^2}{[\boldsymbol{k}_L^2-m_k^2][\boldsymbol{p}_L^2-m_f^2][\boldsymbol{q}_L^2-m_f^2]},\label{G_ini}
\end{align}
\end{widetext}
and $\boldsymbol{k}_L=(i\omega_\ell^B,k_z)$, $\boldsymbol{p}_L=(i\omega_{n_1}^F,p_z)$, $\boldsymbol{q}_L=(i\omega_{n_2}^F,q_z)$. Here we restrict our discussion to the case where $\mu=0$. At this point one can follow two different paths:
\begin{itemize}
\item First evaluate the Matsubara sums and then the momentum integrations. This was the path followed in Ref. \cite{Blaizot:2012sd}. This has the advantage of producing an expression that is also valid in the case where $\mu\ne 0$. However, the resulting integrals are quite involved numerically due to intertwined divergences.

\item First evaluate the momentum integrals and then carry out the Matsubara sums numerically at $\mu =0$. 
This produces a term that depends on temperature and magnetic field which can not be separated into vacuum and medium contributions. 
This is the path we will fallow in this work.
\end{itemize}

Using the Dirac delta and the Kronecker delta in Eq. (\ref{G_ini}), one obtains
\begin{widetext}
\begin{align}
 \mathcal{G}(m_k^2,m_f^2)=&-\beta V g^2 \left(\frac{N_c^2-1}{2}\right)\int\frac{dp_zdq_z}{(2\pi)^2}T^2\sum_{\ell,n_2}\frac{4m_f^2}{[\omega_\ell^2+(p_z-q_z)^2+m_k^2][(\omega_{n_2}+\omega_\ell)^2+p_z^2+m_f^2][\omega_{n_2}^2+q_z^2+m_f^2]} \, ,
\end{align}
\end{widetext}
where $\omega_\ell = 2\pi \ell T$ and $\omega_{n_2}=(2n_2+1)\pi T$.
Now one can first compute the integrals, which yields
\begin{align}
\begin{split}
\mathcal{G}(m_k^2,m_f^2)=&-\beta V g^2\left(\frac{N_c^2-1}{2}\right)T^2 m_f^2
\sum_{\ell,n_2}\frac{\mathcal{E}_\ell-\mathcal{E}_{n_2}}{\mathcal{E}_\ell \mathcal{E}_{n_1} \mathcal{E}_{n_2} \left|\mathcal{E}_\ell-\mathcal{E}_{n_2}\right|\left(\left|\mathcal{E}_\ell-\mathcal{E}_{n_2}\right|+\mathcal{E}_{n_1}\right)} \, ,
\end{split}
\end{align}
where $\mathcal{E}_\ell=\sqrt{\omega_\ell^2+m_k^2}$, $\mathcal{E}_{n_1}=\sqrt{(\omega_{n_2}+\omega_\ell)^2+m_f^2}$, and $\mathcal{E}_{n_2}=\sqrt{\omega_{n_2}^2+m_f^2}$.
Then, in Eq. (\ref{P_exch_ini}), for each value of the Matsubara frequencies, one must perform the integrals in $k$. One can use polar coordinates, so that the final expression for the exchange pressure has the form
\begin{align}
\begin{split}
\frac{P_{\rm exch}^{\rm LLL}}{N_c}=&\frac{1}{2}g^2 \left(\frac{N_c^2-1}{2N_c}\right)T^2 \sum_f m_f^2\left(\frac{q_fB}{2\pi}\right)\sum_{\ell,n_2}\int \frac{dm_k}{2\pi}m_k
 e^{-\frac{m_k^2}{2 q_f B}}\frac{\mathcal{E}_\ell-\mathcal{E}_{n_2}}{\mathcal{E}_\ell \mathcal{E}_{n_1} \mathcal{E}_{n_2} \left|\mathcal{E}_\ell-\mathcal{E}_{n_2}\right| \left(\left|\mathcal{E}_\ell-\mathcal{E}_{n_2}\right|+\mathcal{E}_{n_1}\right)} \, .
\end{split}\label{Pexch}
\end{align}
This expression has the advantage of being numerically simple. Its downside, however, is that it only holds for $\mu=0$ and can not be used for cold and dense QCD. Eq. (\ref{Pexch}) is numerically equivalent to that of Ref. \cite{Blaizot:2012sd}. From a simple analysis of Eq. (\ref{Pexch}), one can check that, for $m_f\to 0$, the exchange contribution to the pressure vanishes. This was also reported in Ref. \cite{Blaizot:2012sd}. Another important advantage of Eq. (\ref{Pexch}) is that it is easy to check that the IR domain for the momenta, $m_k\to 0$, is regulated by the fermionic mass and Matsubara frequencies.

Taking into account the 2L contribution from the gluons, given by the well-known formula \cite{Kapusta:2006pm}:
\begin{equation}
P_{\rm 2}^G=-N_c (N_c^2-1)\frac{g^2 T^4}{144} \,,
\end{equation}
the total 2L pressure can be written as:
\begin{equation}
P_{\rm 2L}= P_{\rm free}^G + P_{\rm 2}^G + P_{\rm free}^{\rm LLL} +  P_{\rm exch}^{\rm LLL}
\,.
\label{PNLO}
\end{equation}

\subsection{Chiral condensate and strange quark number susceptibility}

The chiral condensate is a very relevant observable in the investigation of the phase diagram for strong interactions. For massless quarks it is the true order parameter for the chiral transition. When one includes light quark masses, however, this is no longer true but its behavior near the transition (or crossover) still exhibits a ``memory" of this feature, with the condensate varying appreciably but not sharply, so that it can be considered a pseudo order parameter for the chiral transition in this case. Of course, our perturbative analysis is reliable only for very large temperatures and even larger magnetic fields, so that it cannot bring information on the region near the phase transition or crossover. Nevertheless, since there are lattice results for high  temperatures and magnetic fields, the comparison of these two first-principle calculations in this region is certainly relevant.

The condensate is obtained from the pressure as a derivative with respect to the quark mass. So, the $f$-flavor condensate is given by
\begin{align}
\begin{split}
 \left\langle\bar\psi_f\psi_f\right\rangle=-\frac{\partial P_f}{\partial m_f}
 =-\frac{\partial P_{\rm free}^{\rm LLL}}{\partial m_f}-\frac{\partial P_{\rm exch}^{\rm LLL}}{\partial m_f} \, .
 \end{split} 
\end{align}
From the expressions obtained in the previous section, we can derive straightforwardly
\begin{align}
 \frac{\partial P_{\rm free}^{\rm LLL}}{\partial m_f}=-N_c m_f \frac{q_fB}{(2\pi)^2}\left[1+
 \ln x_f+\int d p_z\frac{2 n_F(E_p)}{E_p}\right]
\end{align}
and
\begin{align}
\begin{split}
\frac{\partial P_{\rm exch}^{\rm LLL}}{\partial m_f}=&-\frac{1}{2}g^2 \left(\frac{N_c^2-1}{2}\right)T^2 \left(\frac{q_fB}{2\pi}\right)\sum_{l,n_2}\int \frac{dm_k}{2\pi}m_ke^{-\frac{m_k^2}{2 q_f B}}\\
&\times\bigg \{\frac{m_f^3\left[\mathcal{E}_{n_2}\left|\mathcal{E}_l-\mathcal{E}_{n_2}\right|-\mathcal{E}_{n_1}\left(\mathcal{E}_l-\mathcal{E}_{n2}\right)\right]}{\mathcal{E}_l \mathcal{E}_{n_1}^2 \mathcal{E}_{n_2}^2 \left(\mathcal{E}_l-\mathcal{E}_{n_2}\right) \left(\left|\mathcal{E}_l-\mathcal{E}_{n_2}\right|+\mathcal{E}_{n_1}\right)^2}\\
 &-\frac{m_f\left(\mathcal{E}_l-\mathcal{E}_{n_2}\right)\left[2\left(\omega_{n_2}+\omega_l\right)^2\omega_{n_2}^2+m_f^2\left(\left(\omega_{n_2}+\omega_l\right)^2+\omega_{n_2}^2\right)\right]}{\mathcal{E}_l \mathcal{E}_{n_1}^3 \mathcal{E}_{n_2}^3 \left|\mathcal{E}_l-\mathcal{E}_{n_2}\right| \left(\left|\mathcal{E}_l-\mathcal{E}_{n_2}\right|+\mathcal{E}_{n_1}\right)} \bigg \} \,,
\end{split}
\end{align}
where $n_F$ is the Fermi-Dirac distribution.

On the lattice, one computes the $f$-flavor renormalized condensate
\begin{align}
  \Sigma_f^{r}(B,T)=\frac{m_f}{m_{\pi}^2 f_{\pi}^2}\left[\left\langle\bar\psi_f\psi_f\right\rangle_{B,T}-\left\langle\bar\psi_f\psi_f\right\rangle_{0,0}\right] \, ,
\end{align}
which eliminates additive and multiplicative divergences. Here, $m_\pi=135$ MeV, $f_\pi=86$ MeV, and $m_f=5$ MeV for the light quarks. To obtain the vacuum condensate, one can not simply take the zero-field limit since we assumed very large fields from the outset \cite{Blaizot:2012sd}.

One usually utilizes the renormalized light quark chiral condensate, built from the sum of the up and down quark contributions, to locate the (pseudo-)critical temperature \cite{DElia:2021yvk}. Since we assume very high magnetic fields and temperatures, with the scale hierarchy given by $m_s \ll T \ll \sqrt{eB}$, this subtraction is negligible in our perturbative calculation. However a direct comparison to the renormalized lattice results must happen in scales not so favorable to pQCD, so that deviations are expected. One should also have in mind that, since the perturbative approach can only capture the behavior of the condensate for large temperatures, it is completely insensitive to features related to the crossover or possible first-order phase transition at high magnetic fields.

A different observable that can also be computed and directly compared to available lattice data is the strange quark number susceptibility
\begin{equation}
\chi^s=\frac{1}{T^2}\frac{\partial^2 P} {\partial\mu_s^2} \,,
\label{chi-s}
\end{equation}
which has previously been computed using hard thermal loop resummation at one-loop order \cite{Karmakar:2020mnj}.
Given the presence of a derivative with respect to the chemical potential, pure vacuum terms are excluded. This presents an advantage when comparing lattice results to pQCD, even if the temperature range in the simulations is still far from optimal for this purpose \cite{Endrodi:2015oba,DElia:2021yvk}.

\subsection{Running coupling and strange quark mass}

The pressure and chiral condensate to 2L for $3$ flavors with physical quark masses depend not only on the temperature and magnetic field, but also on the renormalization subtraction point $\bar{\Lambda}$, an additional mass scale generated by the perturbative expansion. This comes about via the scale dependence of both the strong coupling $\alpha_s(\bar{\Lambda})$ and strange quark masses $m_s(\bar{\Lambda})$. 

The running of both $\alpha_s$ and $m_s$ are known to four-loop order in the $\overline{\rm MS}$ scheme \cite{Vermaseren:1997fq}. Since we have determined the pressure and chiral condensate only to first order in $\alpha_s$, we use for the coupling \cite{Fraga:2004gz}
     \begin{equation}
     \alpha_{s}(\bar{\Lambda})=\frac{4\pi}{\beta_{0}L}\left(
     1-\frac{2\beta_{1}}{\beta^{2}_{0}}\frac{\ln{L}}{L}\right) \,,
     \label{eq:alphas}
     \end{equation}
where $\beta_{0}=11-2N_{f}/3$, $\beta_{1}=51-19N_{f}/3$, $L=2\ln\left(\bar{\Lambda}/\Lambda_{\rm \overline{MS}}\right)$. Since $\alpha_{s}$ depends on $N_{f}$, fixing the massive quark at some energy scale also depends on the number of flavors.
For the strange quark mass, we have
	\begin{eqnarray}
	\begin{aligned}
	m_{s}(\bar{\Lambda})=\hat{m}_{s}\left(\frac{\alpha_{s}}{\pi}\right)^{4/9}
	\left[1+0.895062\left(\frac{\alpha_{s}}{\pi}\right)  
 \right] \;,
	\end{aligned}
	\label{eq:smass}
	\end{eqnarray}
with $\hat{m}_{s}$ being the renormalization group invariant strange quark mass, i.e. $\bar{\Lambda}$ independent. Since Eq. (\ref{eq:alphas}) for $\alpha_{s}$ tells us that different values of $N_{f}$ give different values of $\Lambda_{\overline{\rm MS}}$, by choosing  $\alpha_{s}(\bar{\Lambda}=1.5~{\rm GeV},~N_{f}=3)=0.336^{+0.012}_{-0.008}$ \cite{Bazavov:2014soa}, we obtain $\Lambda^{2+1}_{\overline{\rm MS}}=343^{+18}_{-12}~$MeV. Fixing the strange quark mass at $m_{s}(2~{\rm GeV}, N_{f}=3)=92.4(1.5)~$MeV \cite{Chakraborty:2014aca} gives $\hat{m}^{2+1}_{s}~{\approx}~248.7~$MeV when using $\alpha^{2+1}_{s}$ in Eq. (\ref{eq:smass}).

As usual, there is arbitrariness in the way one should connect the renormalization scale $\bar{\Lambda}$ to a physical mass scale of the system under consideration \cite{Kapusta:2006pm}. In thermal QCD where, besides quark masses, the only scale is given by the temperature, and $T\gg m_f$, the usual choice is the Matsubara frequency $2\pi T$ with a band around it, i.e. $\pi T < \bar{\Lambda} < 4\pi T$. In the present case, where the magnetic field also provides a relevant mass scale given by $\sqrt{eB}$, the choice becomes more ambiguous. Therefore, in the literature of thermal magnetic QCD, one can find a few different assumptions for the form of the running coupling. 

\begin{figure*}[!ht]
\begin{subfigure}
 \centering
 \includegraphics[width=0.45\textwidth]{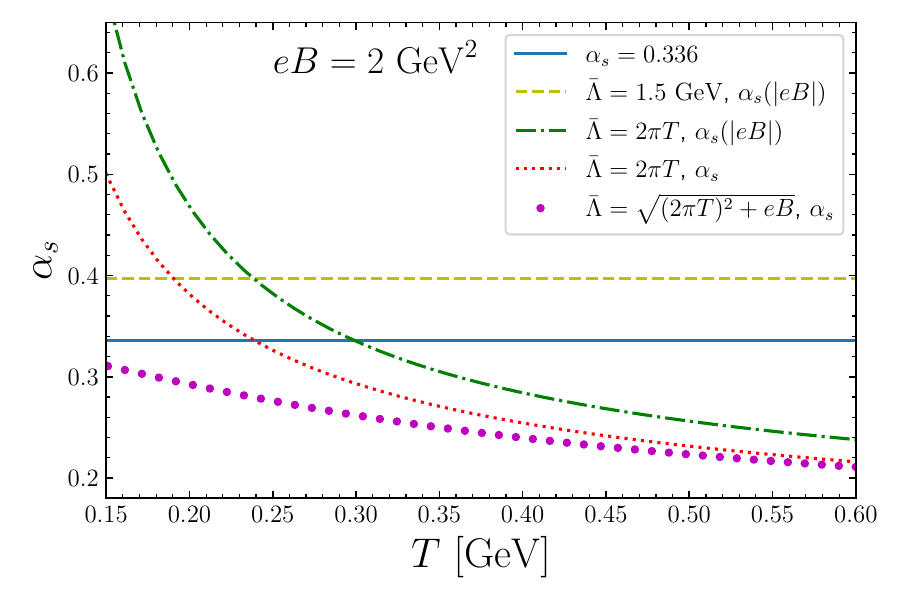} 
 \end{subfigure}
 \begin{subfigure}
  \centering
 \includegraphics[width=0.45\textwidth]{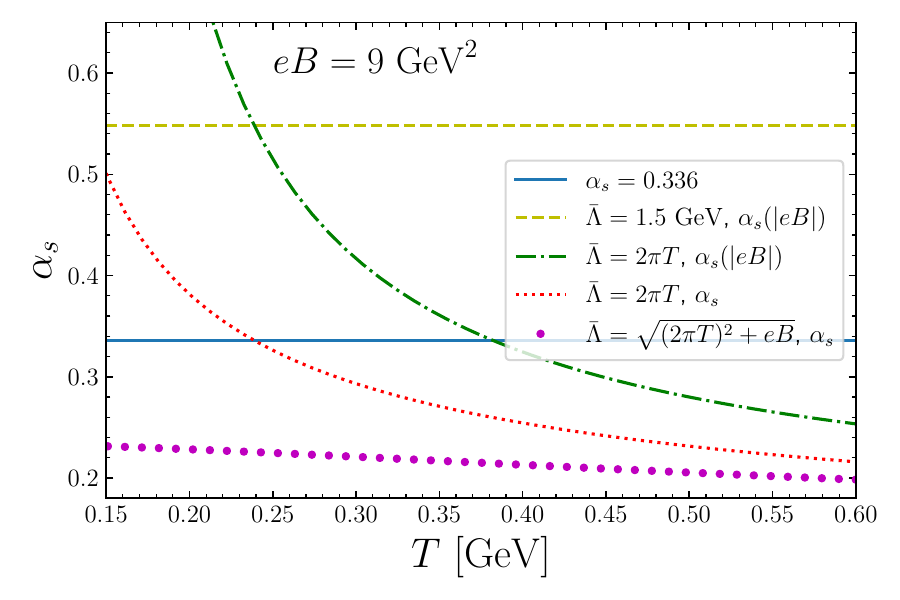}
 \end{subfigure}
\begin{subfigure}
  \centering
 \includegraphics[width=0.45\textwidth]{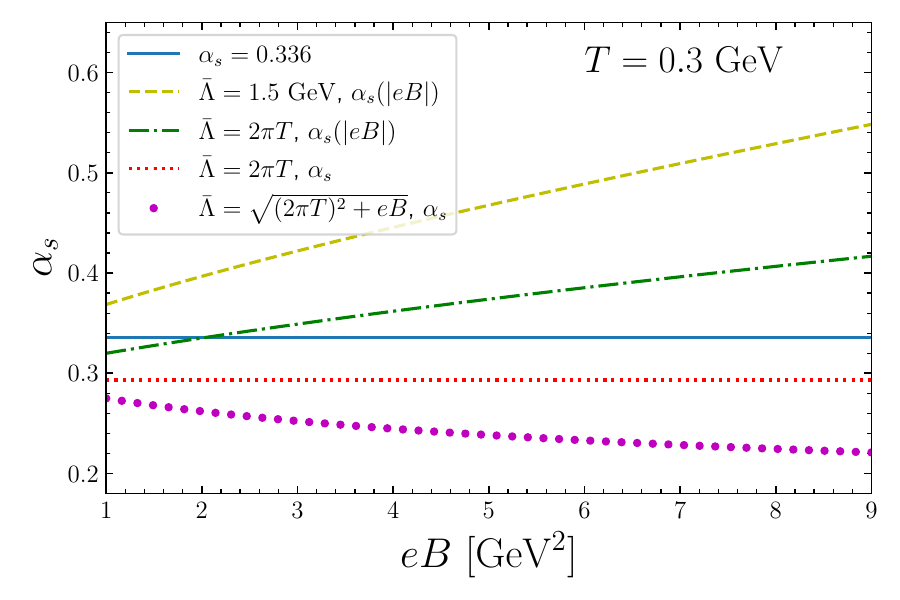}
 \end{subfigure}
 \begin{subfigure}
  \centering
 \includegraphics[width=0.45\textwidth]{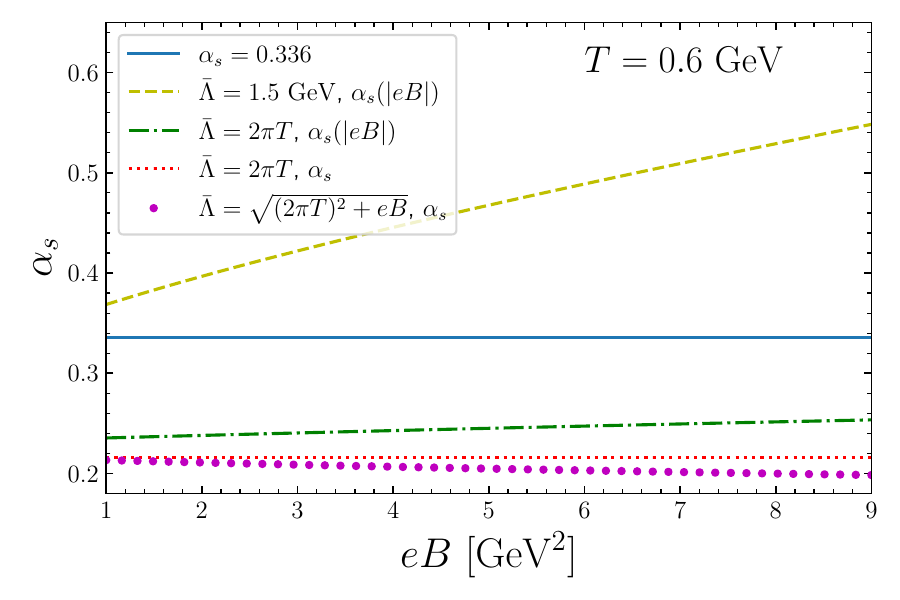}
 \end{subfigure}
\caption{Strong coupling, $\alpha_s$, as a function of temperature for two different (large) values of the magnetic field (top) and as a function of magnetic field at two different temperatures (bottom). For comparison, we show curves for the different choices of the running of $\alpha_s$ discussed in the text.}
\label{fig:alphavsT_B}
\end{figure*}

Since this issue has induced some debate, we show results for a few representative choices and discuss their implications for our observables. Although we have our preference for the most physical choice, we believe that, ultimately, this  will be settled by direct comparison to lattice QCD simulations. Since this problem will also arise in a realm of parameter space still unreachable by Monte Carlo methods, due to the Sign Problem, understanding this in thermal magnetic QCD becomes even more relevant. In what follows, we show results for the following cases:
\begin{itemize}
    \item[(i)] A fixed value of $\alpha_s=0.336$. This corresponds, essentially, to ignoring all the effects from the renormalization group running.
    
    \item[(ii)] The running form proposed in Ref. \cite{Ayala:2018wux}:
\begin{align}
\alpha_s(|eB|)=\frac{\overline{\alpha}_s(\bar{\Lambda}^2)}{1+(\beta_0/4\pi) \overline{\alpha}_s(\bar{\Lambda}^2) \ln(\frac{\bar{\Lambda}^2}{\bar{\Lambda}^2+|eB|})} \; ,
\label{alpha_ayala}
\end{align}
where $\overline{\alpha}_s(\bar{\Lambda}^2)$ corresponds to the usual $\overline{\rm{MS}}$ one-loop 
running 
coupling. Here $\bar{\Lambda}=1.5$ GeV.
The main motivation in this reference has been to try to provide an understanding of the phenomenon of inverse magnetic catalysis (for a review, cf. Ref. \cite{Shovkovy:2012zn}). As will be clear below, however, this form for the running coupling displays an odd behavior as one plays with the magnetic field strength.

\item[(iii)] Same as the previous one, but with $\bar{\Lambda}=2\pi T$. This choice has been adopted, e.g., in Ref. \cite{Bandyopadhyay:2017cle,Karmakar:2019tdp}.
\item[(iv)] $\alpha_s$ given by Eq. (\ref{eq:alphas}) and $\bar{\Lambda}=2\pi T$. This corresponds to the usual thermal QCD choice, and ignores the possible effect of the magnetic field on the scale $\bar{\Lambda}$. 

\item[(v)] Same as the previous one, but with $\bar{\Lambda}=\sqrt{(2\pi T)^2+eB}$. This is, in our view, the most natural and physical choice, which is an extension of what is done in finite-temperature field theory \cite{Kapusta:2006pm}.

\end{itemize}

The running of the strange quark mass will, obviously, be affected by the choice for the running of $\alpha_s$.

\begin{figure*}[!ht]
\begin{subfigure}
 \centering
 \includegraphics[width=0.45\textwidth]{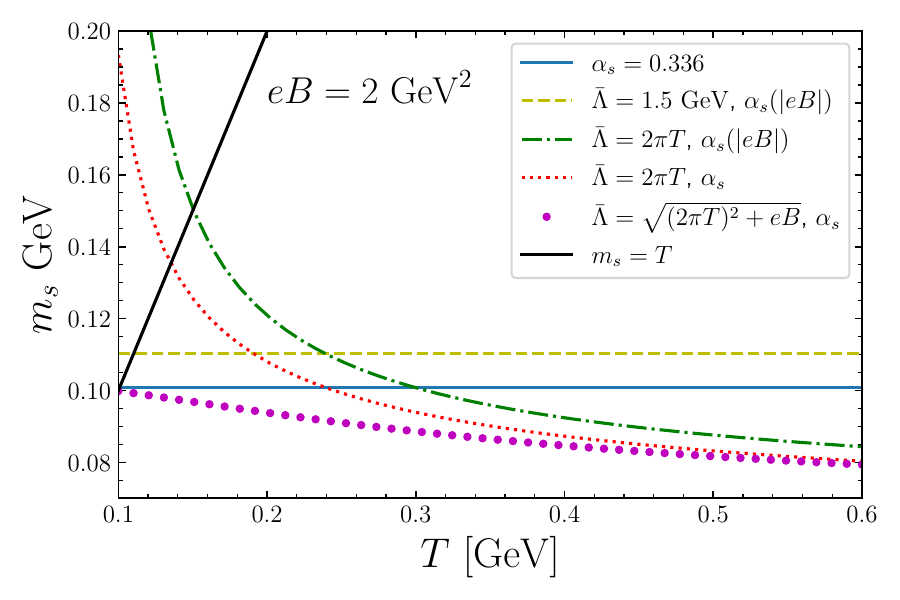} 
 \end{subfigure}
 \begin{subfigure}
  \centering
 \includegraphics[width=0.45\textwidth]{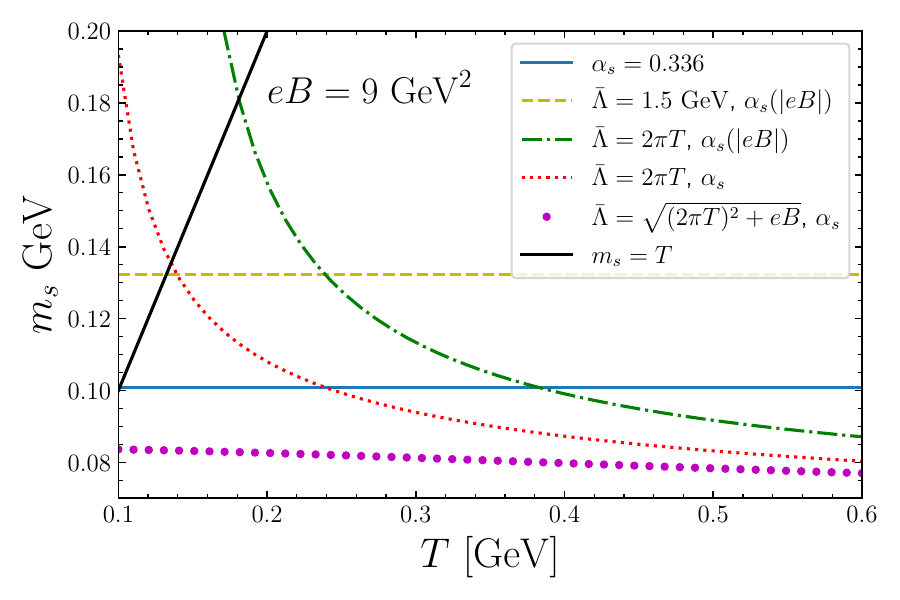}
 \end{subfigure}
\begin{subfigure}
  \centering
 \includegraphics[width=0.45\textwidth]{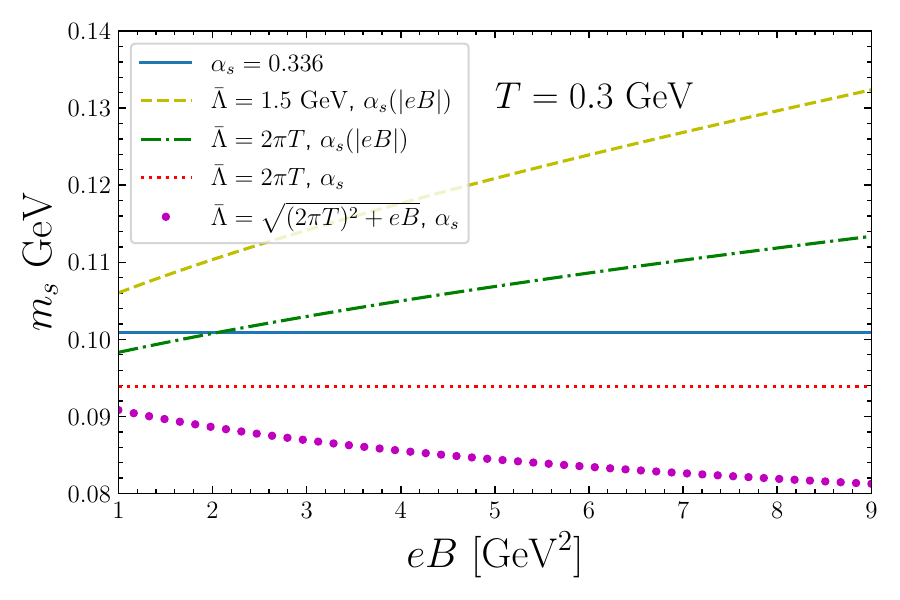}
 \end{subfigure}
 \begin{subfigure}
  \centering
 \includegraphics[width=0.45\textwidth]{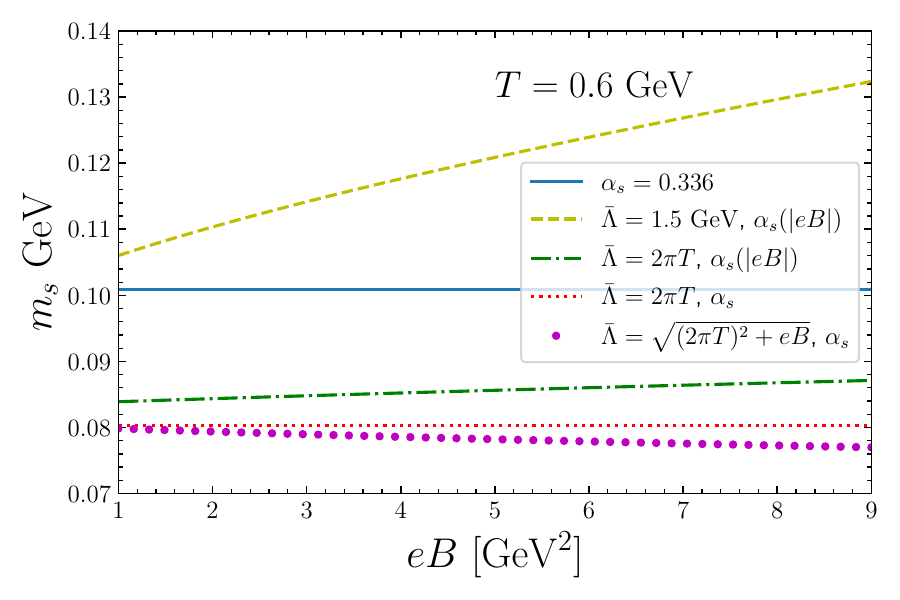}
 \end{subfigure}
\caption{Strange quark mass, $m_s$, as function of temperature for two different (large) values of the magnetic field (top) and as function of magnetic field at two different temperatures (bottom). For comparison, we show curves for the different choices of the running of $\alpha_s$ discussed in the text. The black continuous line corresponds to $m_s=T$ and is there as a reminder of the constraint $m_s\ll T$.}
\label{fig:msvsT_B}
\end{figure*}

In Fig. \ref{fig:alphavsT_B} we show the running of $\alpha_s$ as a function of temperature for two different (large) values of the magnetic field and as a function of the magnetic field strength for two different temperatures. Temperatures are chosen to be large, since we are using perturbative QCD, but within the region of validity for the use of the lowest Landau level approximation, as discussed previously. We also include the case without running ($\alpha_s = 0.336$, case (i)) which provides a scale for comparison. One can verify that cases (ii) and (iii) display a possibly unphysical behavior with increasing magnetic field, since $\alpha_s$ simply grows while the energy density is also increasing. First, it renders perturbative calculations meaningless for high magnetic fields. Second,
it seems incompatible with the expected asymptotic freedom property of strong interactions. In cases (iv) and (v), $\alpha_s$ exhibits the same qualitative (usual) behavior. The quantitative difference comes about because in case (v) the magnetic field contributes to the running scale on an equal footing with respect to the temperature.

In Fig. \ref{fig:msvsT_B} we show the running of the strange quark mass, $m_s$, as a function of temperature for two different (large) values of the magnetic field and as a function of the magnetic field strength for two different temperatures. Temperatures are again chosen to be large, since we are using perturbative QCD, but within the region of validity for the use of the lowest Landau level approximation. We included a black continuous line for $m_s=T$ as a reminder that one has the constraint $m_s\ll T$. The behavior of the different running cases is analogous to what has been discussed for Fig. \ref{fig:alphavsT_B}. The fact that the quark mass increases with magnetic field is probably related to the original motivation of running choices like cases (ii) and (iii), namely, trying to encode magnetic catalysis and inverse magnetic catalysis in the properties of the running of the strong coupling \cite{Ayala:2015bgv,Ayala:2018wux}.

From the discussion above, we believe that only cases (iv) or (v) could be regarded as providing a physical description of the running coupling and running quark mass. Nevertheless, since it can also be tested by direct comparison to lattice data, we will keep all cases in our results for the pressure, chiral condensate and strange quark number susceptibility.

\begin{figure*}[!ht]
\begin{subfigure}
 \centering
 \includegraphics[width=0.45\textwidth]{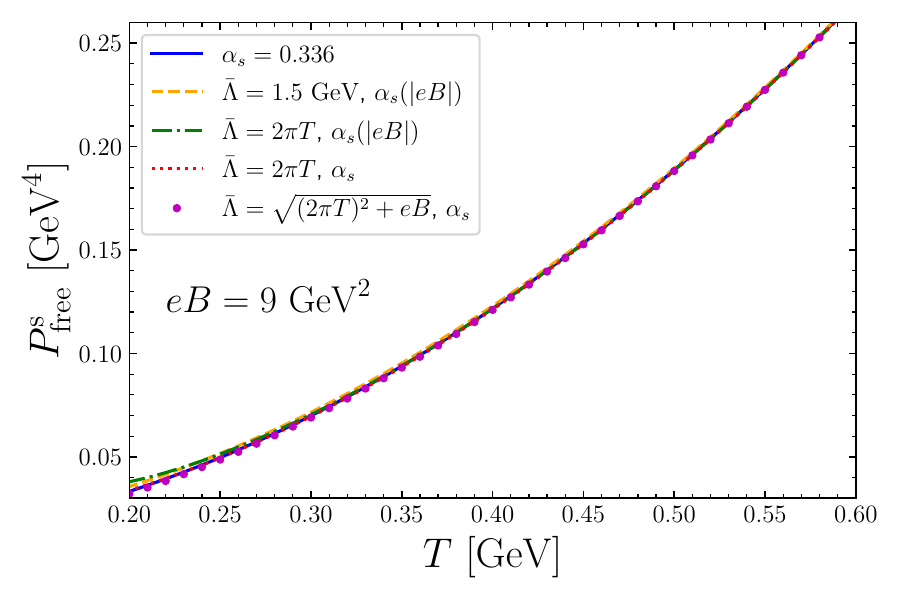} 
 \end{subfigure}
 \begin{subfigure}
  \centering
 \includegraphics[width=0.45\textwidth]{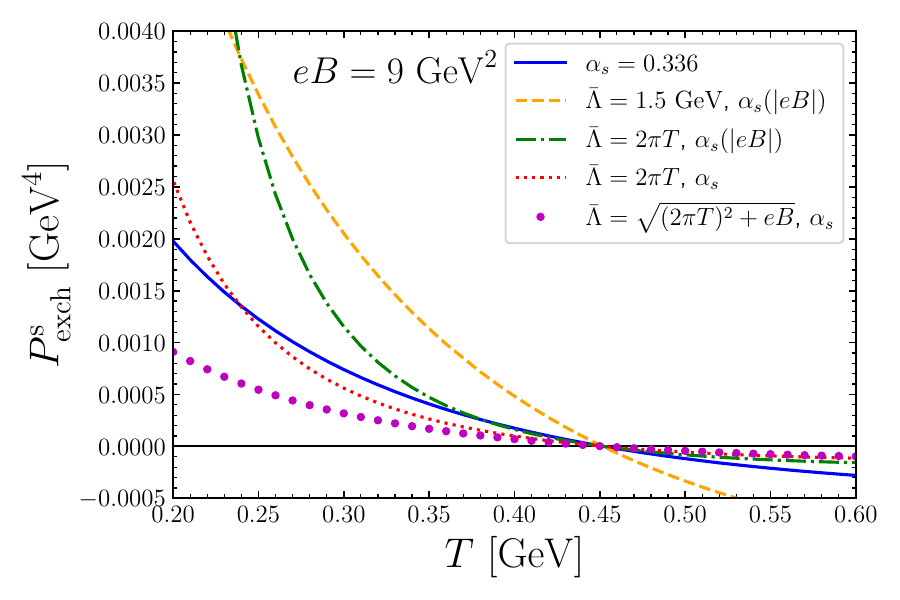}
 \end{subfigure}
\begin{subfigure}
  \centering
 \includegraphics[width=0.45\textwidth]{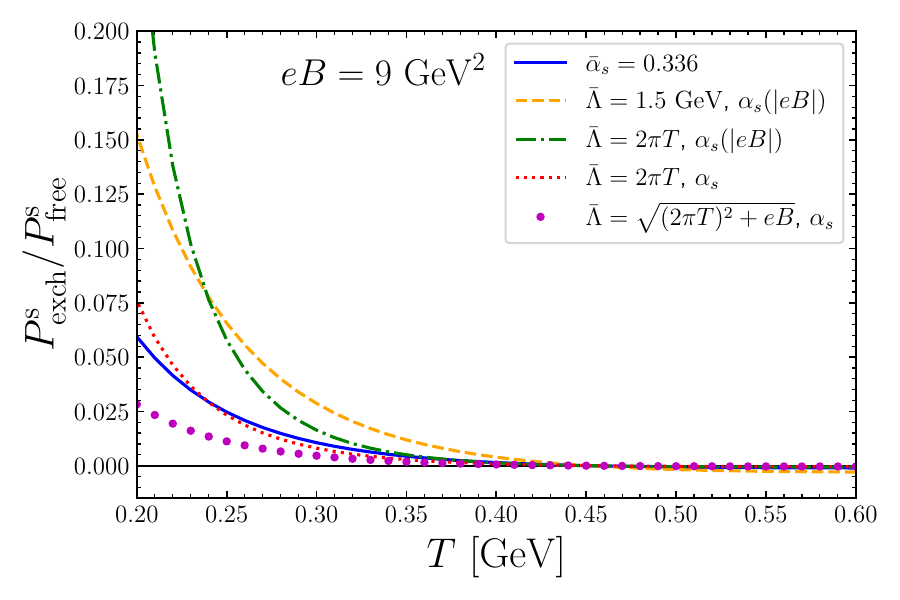}
 \end{subfigure}
 \begin{subfigure}
 \centering
 \includegraphics[width=0.45\textwidth]{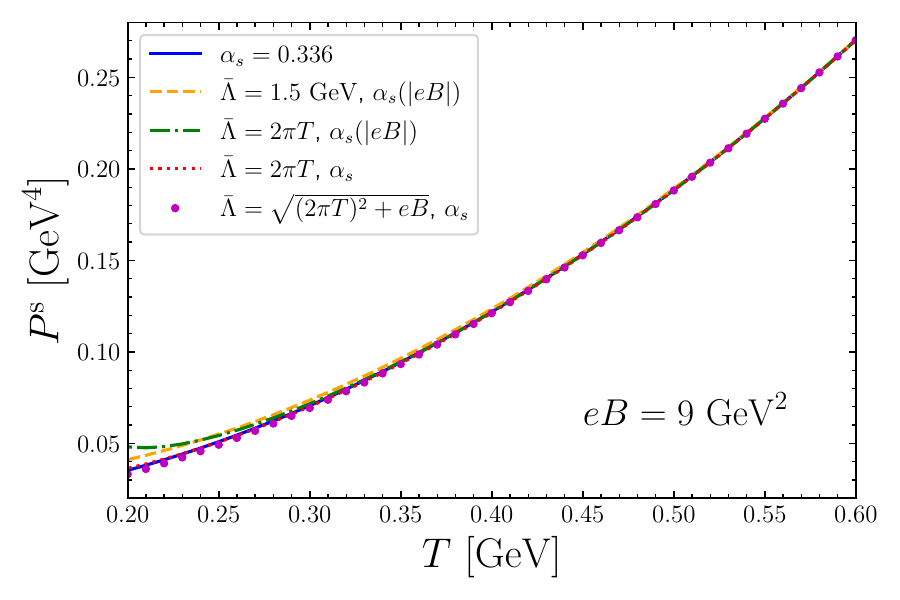} 
 \end{subfigure}
\caption{$P^s_{\rm free}$ (top-left), $P^s_{\rm exch}$ (top-right), $P^s_{\rm exch}/P^s_{\rm free}$ (bottom-left), and $P^s$ (bottom-right) as functions of the temperature for $eB=9$ $\rm{GeV}^2$.}
\label{fig:Ps_B9}
\end{figure*}

\section{Results}
\label{sec:results}

We can now discuss our perturbative results for the pressure, chiral condensate and strange quark number susceptibility to 2L for very large magnetic fields. We show results for the different running schemes discussed in the literature, and compare them to what has been obtained recently on the lattice. 

\subsection{Pressure}

\begin{figure*}[!ht]
\begin{subfigure}
 \centering
 \includegraphics[width=0.45\textwidth]{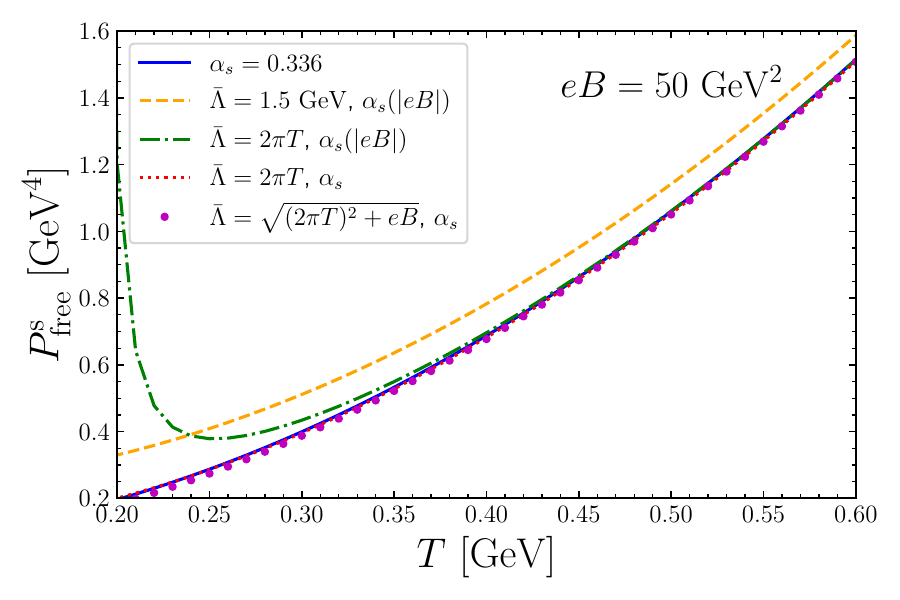} 
 \end{subfigure}
 \begin{subfigure}
  \centering
 \includegraphics[width=0.45\textwidth]{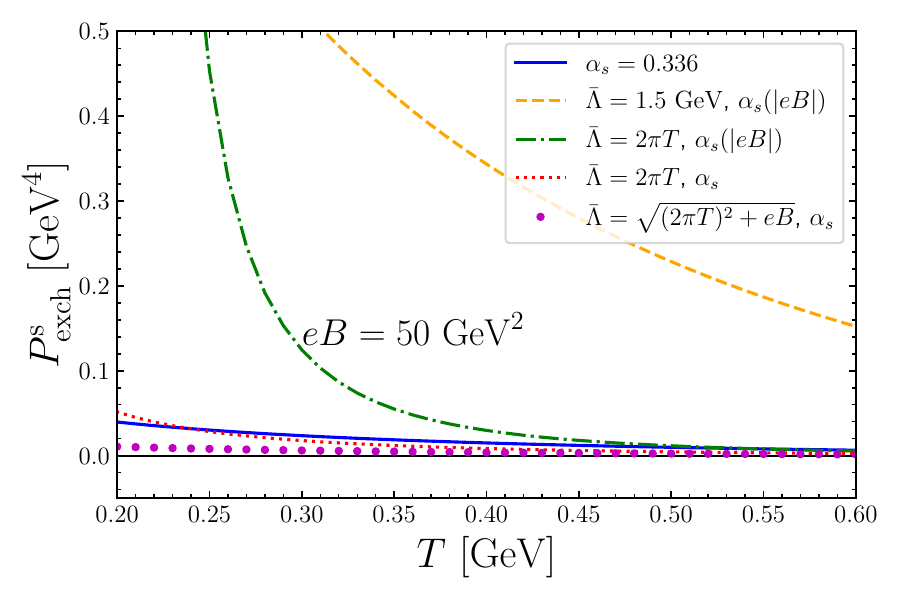}
 \end{subfigure}
\begin{subfigure}
  \centering
 \includegraphics[width=0.45\textwidth]{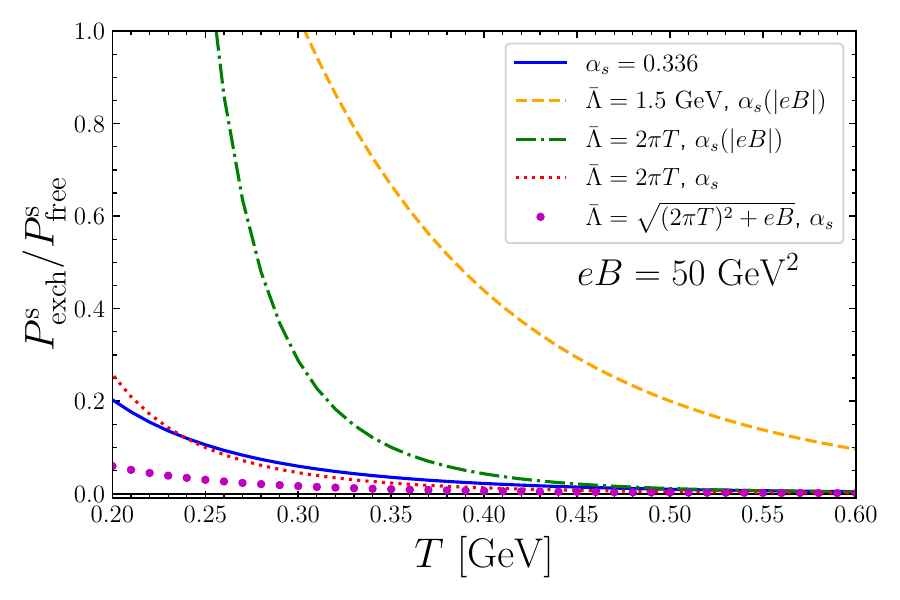}
 \end{subfigure}
 \begin{subfigure}
 \centering
 \includegraphics[width=0.45\textwidth]{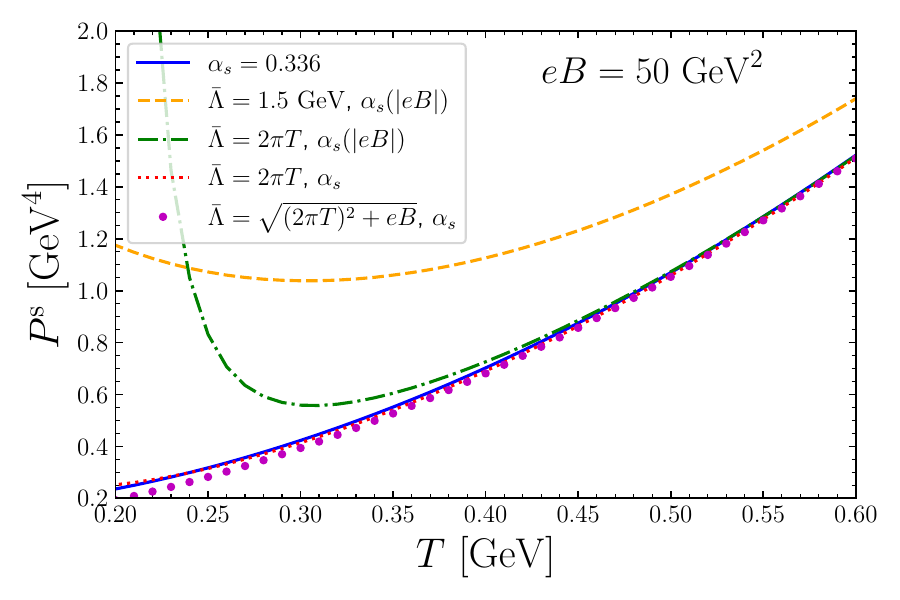} 
 \end{subfigure}
\caption{$P^s_{\rm free}$ (top-left), $P^s_{\rm exch}$ (top-right), $P^s_{\rm exch}/P^s_{\rm free}$ (bottom-left), and $P^s$ (bottom-right) as functions of the temperature for $eB=50$ $\rm{GeV}^2$.}
\label{fig:Ps_B50}
\end{figure*}

\begin{figure*}[!ht]
\begin{subfigure}
 \centering
 \includegraphics[width=0.45\textwidth]{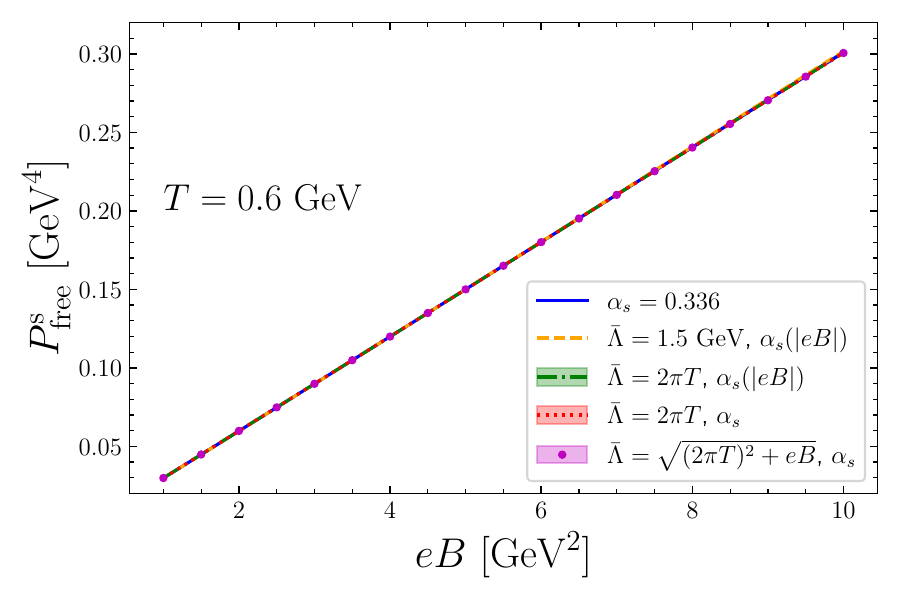} 
 \end{subfigure}
 \begin{subfigure}
  \centering
 \includegraphics[width=0.45\textwidth]{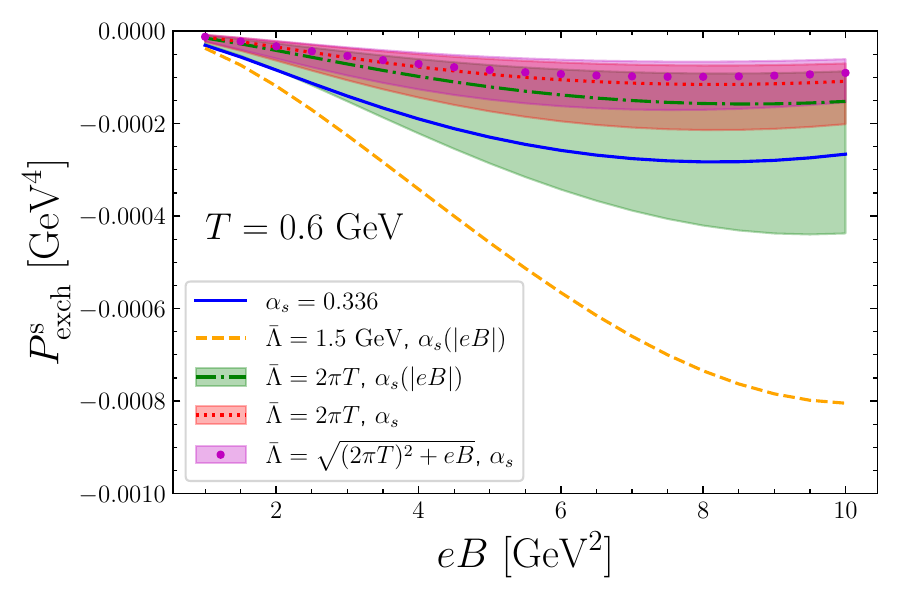}
 \end{subfigure}
\begin{subfigure}
  \centering
 \includegraphics[width=0.45\textwidth]{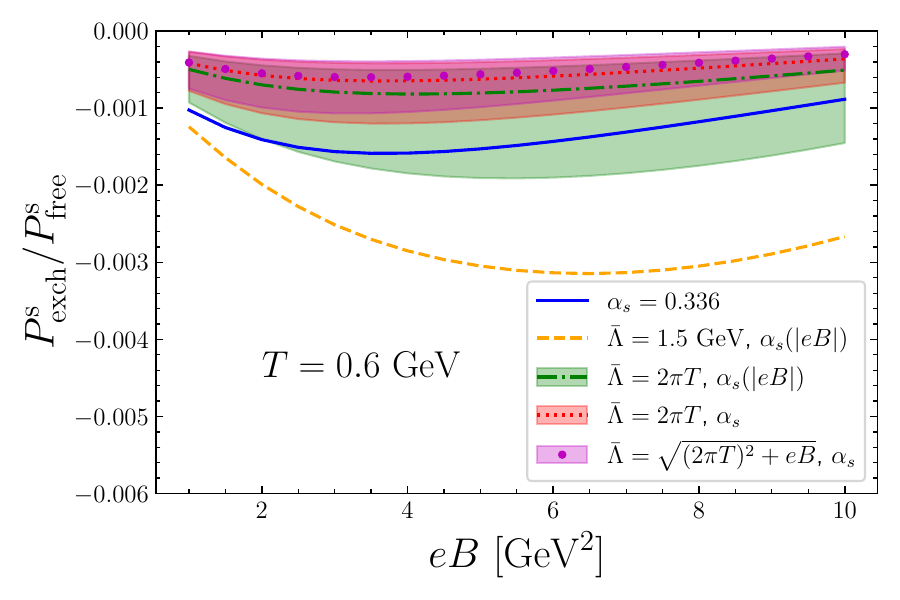}
 \end{subfigure}
 \begin{subfigure}
 \centering
 \includegraphics[width=0.45\textwidth]{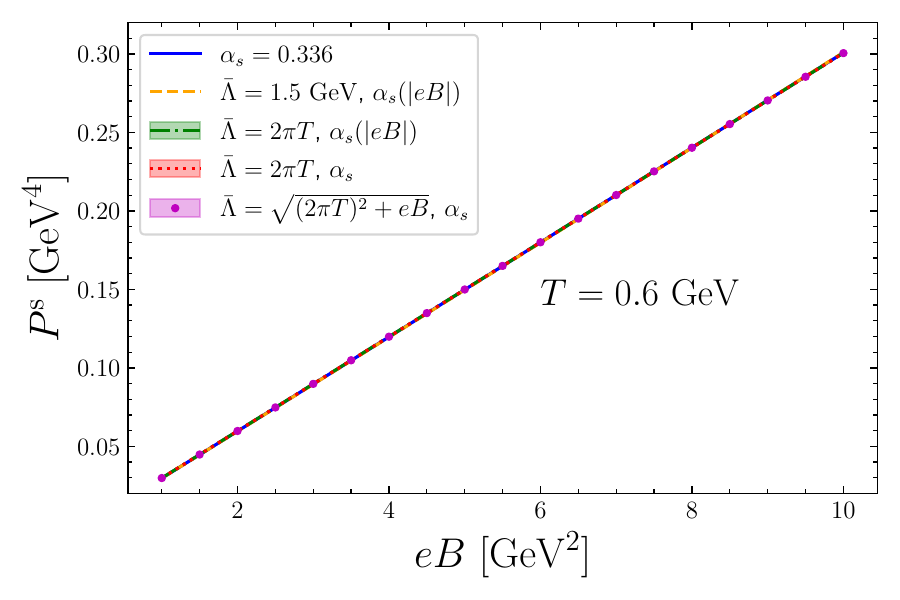} 
 \end{subfigure}
\caption{$P^s_{\rm free}$ (top-left), $P^s_{\rm exch}$ (top-right), $P^s_{\rm exch}/P^s_{\rm free}$ (bottom-left), and $P^s$ (bottom-right) as functions of the magnetic field at $T=0.6$ $\rm{GeV}$. The bands correspond to changes in the central scale by a factor of $2$.}
\label{fig:Ps_T6_bands}
\end{figure*}

In what follows, we present results for the pressure as a function of the temperature for the highest value of magnetic field attained in present lattice simulations ($eB=9$ GeV$^2$), Fig. \ref{fig:Ps_B9}, and for an even larger field ($eB=50$ GeV$^2$), Fig. \ref{fig:Ps_B50}. We also present results for the pressure as a function of the the magnetic field for $T=0.6$ GeV (Fig. \ref{fig:Ps_T6_bands}). In these figures, we show a panel with the free pressure, $P^s_{\rm free}$, the exchange diagram contribution, $P^s_{\rm exch}$, the ratio $P^s_{\rm exch}/P^s_{\rm free}$, and the full strange pressure, $P^s$. We show results for the contribution from the strange quark because mass effects are more relevant in this case. The ratio $P^s_{\rm exch}/P^s_{\rm free}$ provides a certain measure of the reliability of perturbation theory, since it seems to be more well behaved than the case in the absence of a large magnetic field \cite{Blaizot:2012sd}.

\begin{figure*}[!ht]
  \centering
 \includegraphics[width=0.45\textwidth]{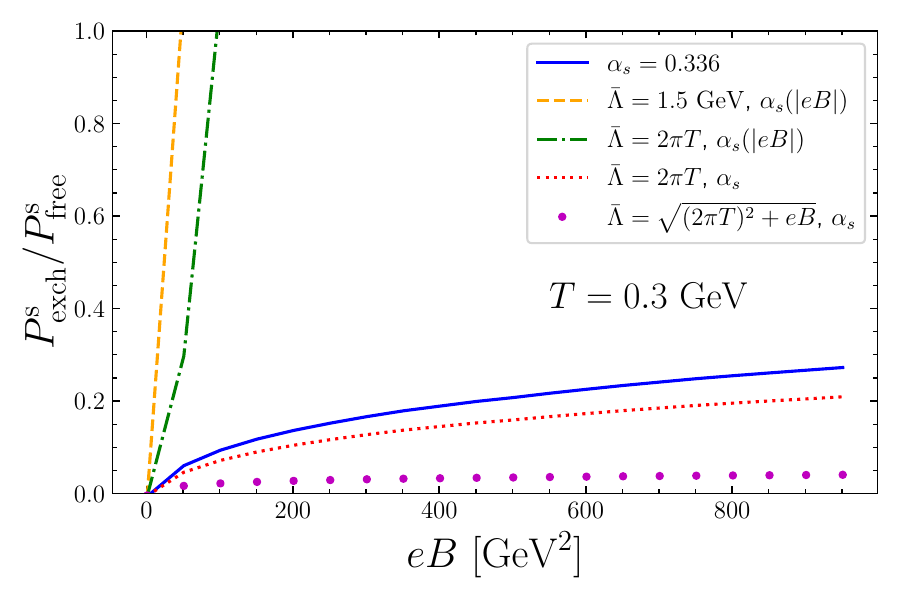}
  \centering
\caption{$P^s_{\rm{exch}}/P^s_{\rm{free}}$ as a function of the magnetic field for $T=0.3$ $\rm{GeV}$.}

\label{fig:Ps_T3_Bext}
\end{figure*}

Finally, for the sake of completeness, we show how the pressure behaves for huge values of the magnetic field $eB=10^3$ GeV$^2$ (Fig. \ref{fig:Ps_T3_Bext}). For such high fields, one should definitely take into account anisotropy effects \cite{Bali:2013esa,Huang2010}, which we fully neglect for simplicity. Results shown here would correspond to the longitudinal pressure in an anisotropic description \cite{Bali:2014kia}. 
For phenomenological applications, one usually has to take into account effects from anisotropy.

In Figs. \ref{fig:Ps_B9} and \ref{fig:Ps_B50} we can observe how the behavior of the pressure is modified for the different choices of the running of the strong coupling. For the cases (ii) and (iii) discussed in the previous section, one finds a much poorer convergence, which becomes worse as one increases the magnetic field. This is compatible with the somewhat unphysical behavior observed in the running of $\alpha_s$ and $m_s$ for these choices of renormalization scale. Cases (iv) and (v), on the other hand, seem to be well behaved. One should notice that a future comparison to lattice results will have to take into account the different  vacuum subtraction schemes adopted in lattice simulations, pQCD calculations and effective models \cite{Fraga:2008qn,Mizher:2010zb,Endrodi:2013cs,Haber:2014ula}.

In Fig. \ref{fig:Ps_T6_bands} we display the same cases as before, but as a function of the magnetic field. We also include bands corresponding to increasing/decreasing the central renormalization running scale by a factor of $2$. As usual, the size of these bands correspond to a rough measure of the theoretical uncertainty of the perturbative series, since it represents the residual renormalization scale dependence \cite{Kapusta:2006pm}. Notice that case (ii) has no band by construction, since $\overline\Lambda$ is fixed. 

\begin{figure}
    \centering
    \includegraphics[width=0.45\textwidth]{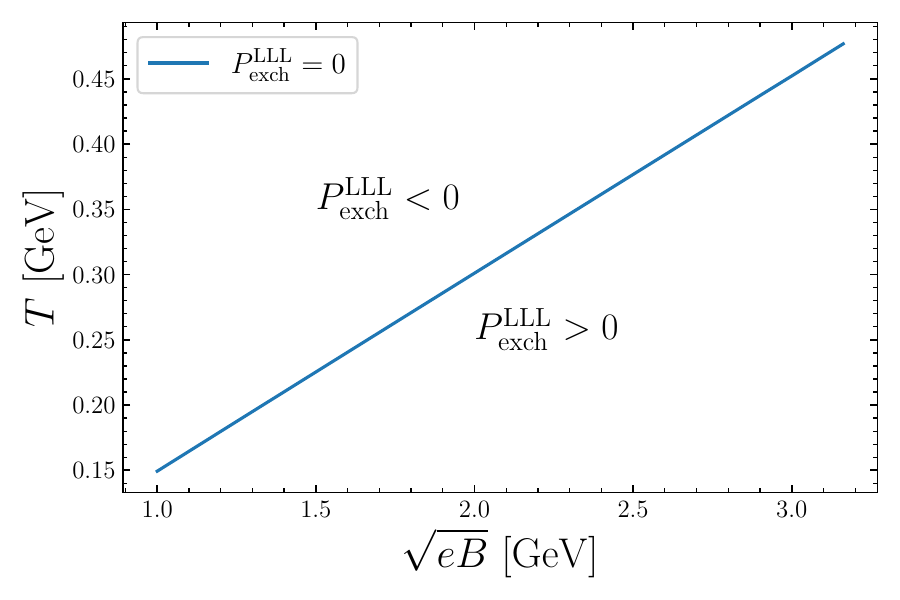}
    \caption{Sign of $P_{\rm{exch}}^{\rm{LLL}}=0$ in the $\sqrt{eB}-T$ plane.}
    \label{fig:P_exch_0}
\end{figure}

From the first and last panel it is clear that the quark pressure is dominated by the free gas  contribution. In Fig. \ref{fig:Ps_T3_Bext}, as we increase the possible values of the external magnetic field dramatically, we see a clear separation in the behavior of cases (ii) and (iii), which are essentially ill defined perturbatively, and cases (iv) and (v), which behave well. Case (i) is trivial, since there is no running. One should notice also that $P_{\rm{exch}}^{\rm{LLL}}$ changes sign depending on the value of the temperature and magnetic field. This behavior is exhibited in Fig. \ref{fig:P_exch_0} in the $\sqrt{eB}-T$ plane. One should emphasize that this behavior is not sensitive to the value of any other parameter.



\begin{figure*}[!ht]
\begin{subfigure}
 \centering
 \includegraphics[width=0.45\textwidth]{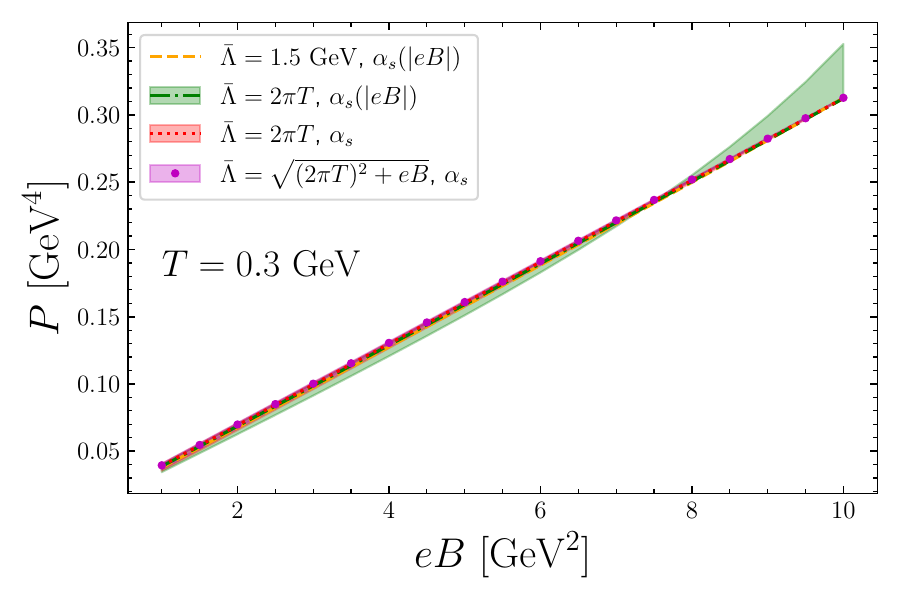} 
 \end{subfigure}
 \begin{subfigure}
  \centering
 \includegraphics[width=0.45\textwidth]{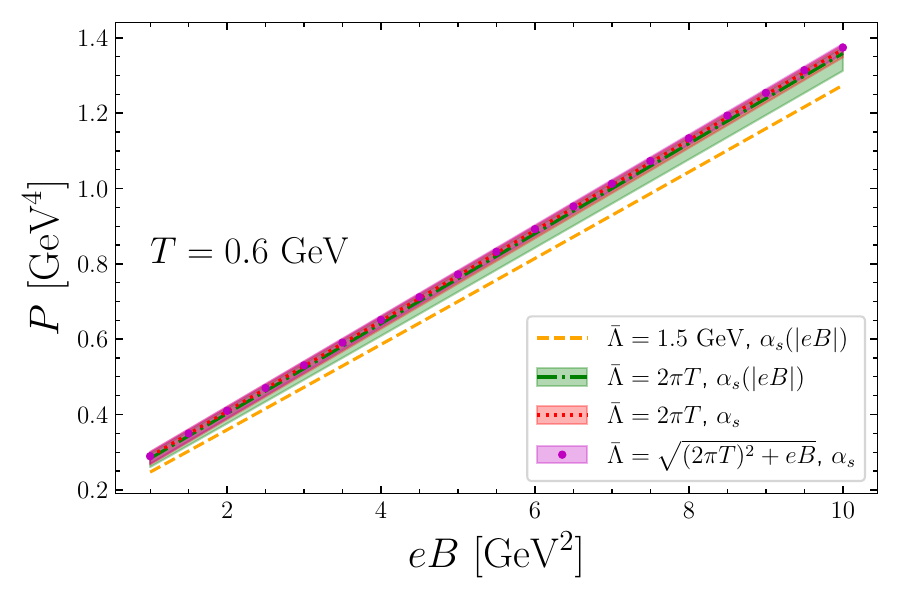}
 \end{subfigure}
\begin{subfigure}
  \centering
 \includegraphics[width=0.45\textwidth]{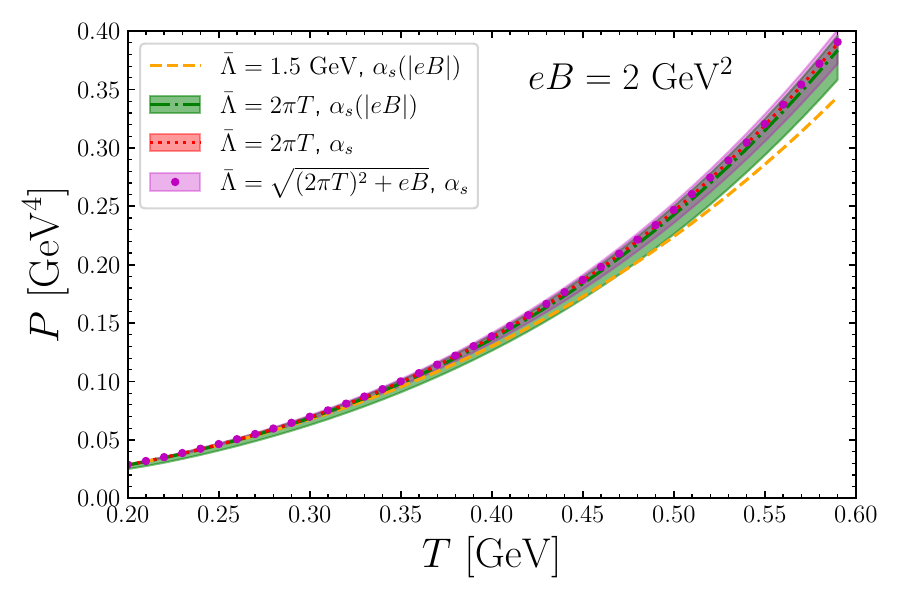}
 \end{subfigure}
 \begin{subfigure}
 \centering
 \includegraphics[width=0.45\textwidth]{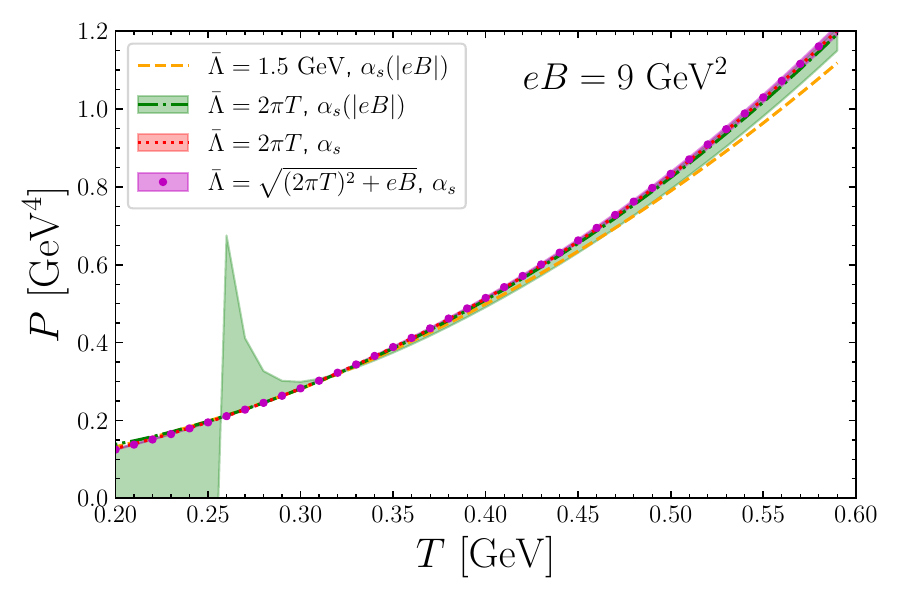} 
 \end{subfigure}
\caption{Full pressure as function of the magnetic filed at $T=0.3$ GeV (top-left) and $T=0.6$ GeV (top-right), and as function of temperature at $eB=2$ $\rm{GeV}^2$ (bottom-left) and $eB=9$ $\rm{GeV}^2$ (bottom-right). The bands correspond to changes in the central scale by a factor of two.}
\label{fig:P_tot_band}
\end{figure*}

The full pressure $P_{\rm 2L}$, given by Eq. (\ref{PNLO}), as a function of the magnetic field for two different values of the temperature, and as a function of the temperature for two different values of the magnetic field, is shown in Fig. \ref{fig:P_tot_band} with bands defined as above. Notice that the green band in the last panel is divergent for small temperatures.

\subsection{Chiral condensate and strange quark number susceptibility}

Now we present our results for the chiral condensate and the strange quark number susceptibility as a function of the temperature in the presence of high magnetic fields. We show results for the highest magnetic fields attained in lattice QCD simulations so far. Of course, the perturbative approach has the caveat of being reliable only for large temperatures, so that we will not be able to describe nontrivial features of the condensate, such as its behavior near the transition (or crossover). In any case, perturbation theory would not be sensitive to such effects.

\begin{figure}[!ht]
\begin{subfigure}
 \centering
 \includegraphics[width=0.45\textwidth]{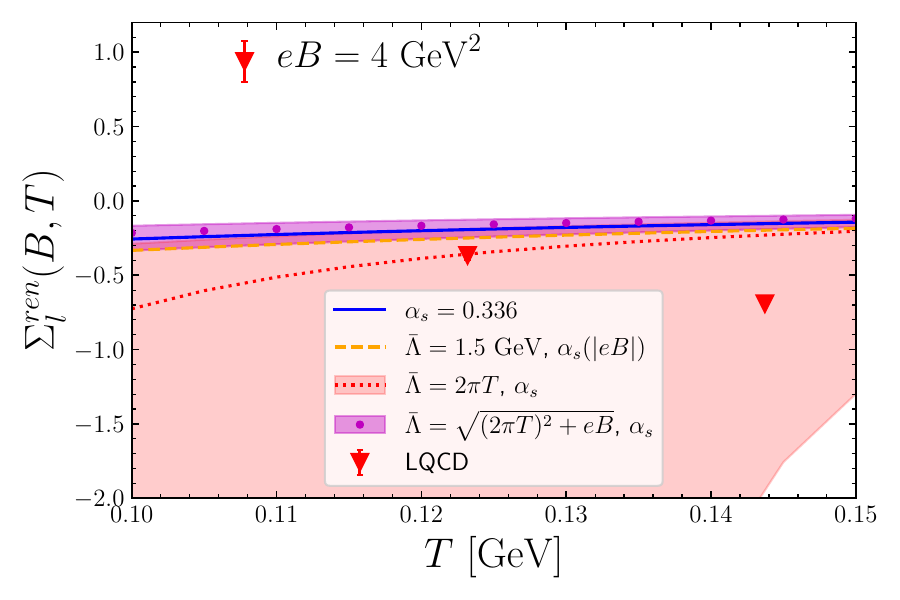} 
 \end{subfigure}
 \begin{subfigure}
  \centering
 \includegraphics[width=0.45\textwidth]{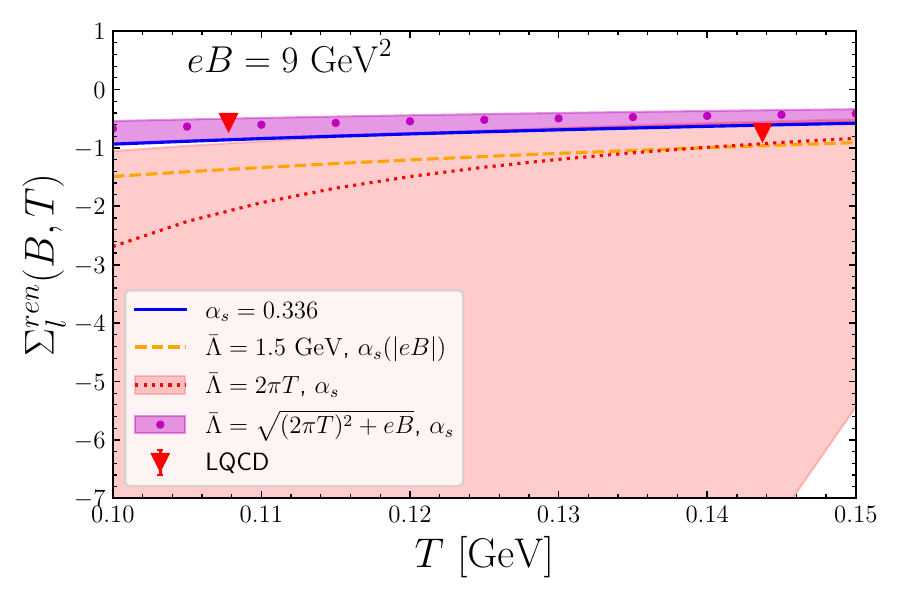}
 \end{subfigure}

\caption{Renormalized light quark chiral condensate as a function of the temperature for $eB=4$ $\rm{GeV}^2$ (left) and $eB=9$ $\rm{GeV}^2$ (right). }
\label{Cond_Sig_B4_9}
\end{figure}

\begin{figure}[!ht]
\begin{subfigure}
 \centering
 \includegraphics[width=0.45\textwidth]{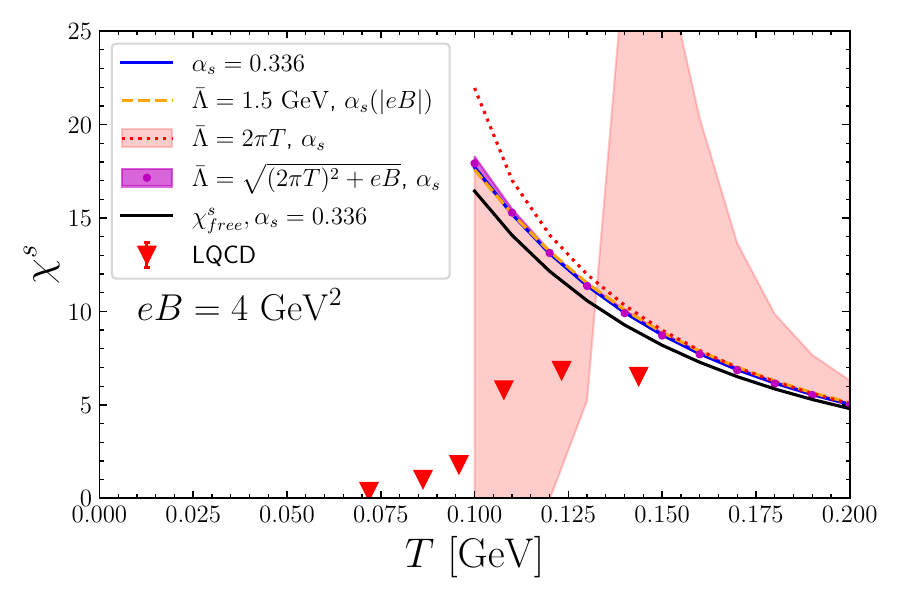} 
 \end{subfigure}
 \begin{subfigure}
  \centering
 \includegraphics[width=0.45\textwidth]{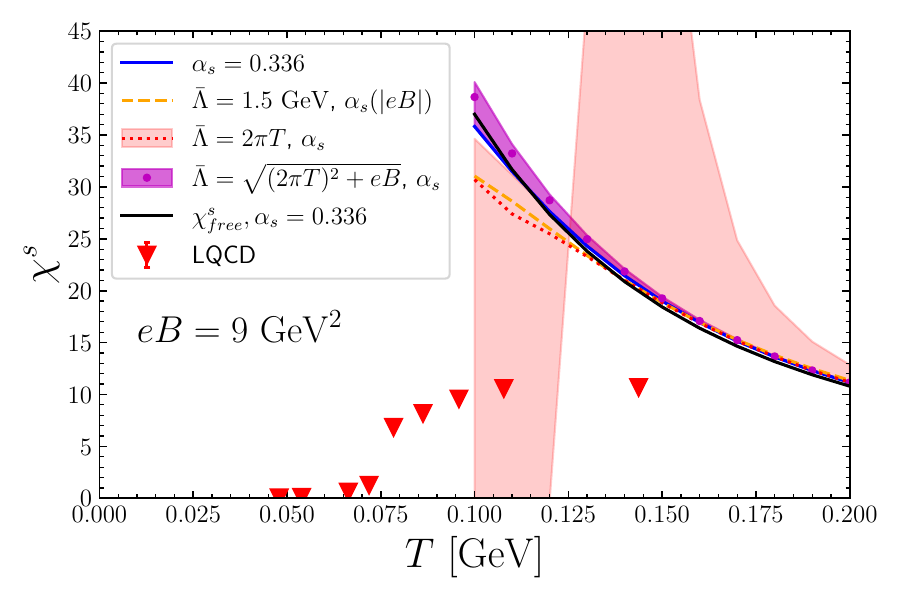}
 \end{subfigure}

\caption{Strange quark number susceptibility as a function of the temperature for $eB=4$ $\rm{GeV}^2$ (left) and $eB=9$ $\rm{GeV}^2$ (right) }
\label{Suscep_B4_9}
\end{figure}

\begin{figure}[!ht]
 \centering
 \includegraphics[width=0.45\textwidth]{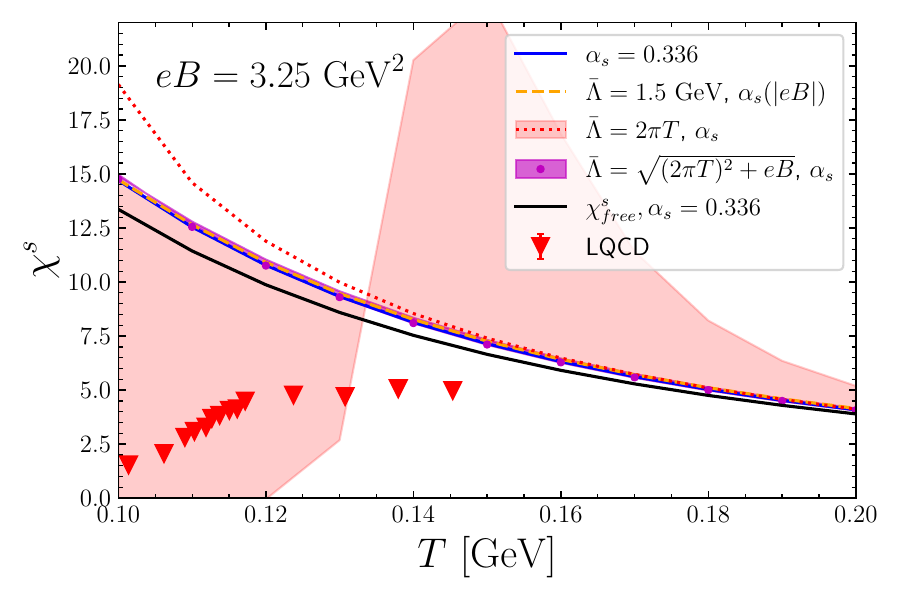} 

\caption{Strange quark number susceptibility  as function of the temperature for $eB=3.25$ }
\label{Suscep_B3_25}
\end{figure}

In Fig. \ref{Cond_Sig_B4_9} we show the renormalized light quark chiral condensate as a function of the temperature for $eB=4$ $\rm{GeV}^2$ and $eB=9$ $\rm{GeV}^2$ computed using perturbative QCD. We also show points obtained via lattice simulations for comparison \cite{DElia:2021yvk}. In Fig. \ref{Suscep_B4_9} we do the same for the strange quark number susceptibility, and, In Fig. \ref{Suscep_B3_25} we show the strange quark number susceptibility for a lower value of magnetic field \cite{Endrodi:2015oba}.

Unfortunately, in both cases the temperature range for lattice results is well below the ideal for a fair comparison to perturbative QCD. Nevertheless, one can see that perturbative results are in the right ballpark for the upper end of temperatures. It is still unclear from the available lattice data whether our calculations capture the qualitative trend at high temperatures. Lattice results for higher temperatures, and even higher magnetic fields, would be necessary for this purpose. As argued previously, the strange quark number susceptibility represents a better observable for our comparison, since in our approach the vacuum contribution is neglected, even though it might still be relevant for the chiral condensate at the temperatures currently accessible to lattice simulations. From the figures one sees that the comparison of pQCD results to lattice data on the strange quark number susceptibility seems to display a more promising trend for temperatures above the ones currently simulated. It is important to note that our framework here is valid only if the hierarchy of scales $m_s \ll T \ll \sqrt{eB}$ is satisfied. In this sense, it is not a problem  if the perturbative results deviate from lattice data for very high temperatures at fixed $eB$.

It is also clear from the plots, moreover, that for such high fields loop corrections to the free case become almost irrelevant, as was already remarked in Ref. \cite{Blaizot:2012sd} in the context of the pressure in the chiral limit. Moreover, the analysis of the different possibilities for the running scale choice show that the width of the band for case (iii) basically diverges (not shown in the figures), case (iv) has a wide band that also diverges at some point for the susceptibility, and case (v) is always well behaved. This behavior is, of course, compatible with what has been observed for the pressure.

\section{Summary and outlook}
\label{sec:outlook}

In this paper we computed the pressure, chiral condensate and strange quark number susceptibility  within perturbative QCD at finite temperature and very high magnetic fields up to two-loop and physical quark masses. Since we adopt the lowest-Landau level approximation in order to obtain analytic results and more control on qualitative aspects, the region of validity for our framework is restricted to $m_s \ll T \ll \sqrt{eB}$, where $m_s$ is the strange quark mass, $e$ is the fundamental electric charge, $T$ is the temperature, and $B$ is the magnetic field strength. 

Since the literature in the field exhibits several possibilities for the running scheme, we study the convergence of the perturbative series\footnote{Of course, one can not assure convergence for higher perturbative orders from our analysis. Nevertheless, for very large magnetic fields perturbation theory seems to be better behaved, as discussed in Ref. \cite{Blaizot:2012sd}.
Our results for the two-loop contribution are very small compared with the one-loop term,  as expected. In any case, here we are only checking convergence when using different running couplings considered in the literature.} for the pressure using the most commonly adopted choices for the scale and functional form of the running coupling, $\alpha_s (T,B)$. Our findings seem to indicate that cases (ii) and (iii) are inconsistent from the point of view of the convergence of the perturbative series, while cases (iv) and (v) pass this criterion, case (v) being the most well behaved. 

Currently, there are essentially two completely opposite scenarios
for the way a magnetic field background affects the QCD interactions: either an enhancement of the strong coupling that renders perturbative calculations not applicable even for physically achievable magnetic fields; or a coupling that is strongly suppressed as the energy density grows, in accordance with usual expectations from asymptotic freedom. It would 
 be desirable to have lattice results that help clarifying this issue.
 
Moreover, the difficulty of choosing a running scale in a setting in which more than one relevant control parameter exists will also be present in the description of systems at finite density. In particular, the physics of magnetars could be sensitive to this choice \cite{Duncan:1992hi,Thompson:1993hn,Kouveliotou:1998ze}. 
 
Our results for the chiral condensate and for the strange quark number susceptibility were directly compared to recent lattice QCD data away from the chiral transition. Even though, as discussed previously, current lattice results do not overlap with the region of validity for our approximations, perturbative results seem to be in the same ballpark, which is encouraging. The window of applicability is still narrow, but our results are obtained from a clean first-principle calculation that can be systematically improved. Furthermore, as argued previously in Ref. \cite{Blaizot:2012sd} for a fixed strong coupling $\alpha_s$, medium loop corrections seem to become essentially negligible as compared to the free term for very high magnetic fields for physical choices of the renormalization running scale. 
\begin{acknowledgments}
We thank M. D'Elia, G. Endrodi and L. Maio for providing lattice data for comparison. We also thank A. Ayala and G. Endrodi for discussions. This work was partially supported by CAPES (Finance Code 001), Conselho Nacional de Desenvolvimento Cient\'{\i}fico e Tecnol\'{o}gico (CNPq), Funda\c c\~ao Carlos Chagas Filho de Amparo \` a Pesquisa do Estado do Rio de Janeiro (FAPERJ), and INCT-FNA (Process No. 464898/2014-5). T.E.R acknowledges support from FAPERJ, Process SEI-260003/002665/2021 and SEI-260003/019683/2022.
\end{acknowledgments}





\bibliographystyle{apsrev4-1}
\bibliography{refs.bib}

\begin{thebibliography}{58}%
\makeatletter
\providecommand \@ifxundefined [1]{%
 \@ifx{#1\undefined}
}%
\providecommand \@ifnum [1]{%
 \ifnum #1\expandafter \@firstoftwo
 \else \expandafter \@secondoftwo
 \fi
}%
\providecommand \@ifx [1]{%
 \ifx #1\expandafter \@firstoftwo
 \else \expandafter \@secondoftwo
 \fi
}%
\providecommand \natexlab [1]{#1}%
\providecommand \enquote  [1]{``#1''}%
\providecommand \bibnamefont  [1]{#1}%
\providecommand \bibfnamefont [1]{#1}%
\providecommand \citenamefont [1]{#1}%
\providecommand \href@noop [0]{\@secondoftwo}%
\providecommand \href [0]{\begingroup \@sanitize@url \@href}%
\providecommand \@href[1]{\@@startlink{#1}\@@href}%
\providecommand \@@href[1]{\endgroup#1\@@endlink}%
\providecommand \@sanitize@url [0]{\catcode `\\12\catcode `\$12\catcode
  `\&12\catcode `\#12\catcode `\^12\catcode `\_12\catcode `\%12\relax}%
\providecommand \@@startlink[1]{}%
\providecommand \@@endlink[0]{}%
\providecommand \url  [0]{\begingroup\@sanitize@url \@url }%
\providecommand \@url [1]{\endgroup\@href {#1}{\urlprefix }}%
\providecommand \urlprefix  [0]{URL }%
\providecommand \Eprint [0]{\href }%
\providecommand \doibase [0]{http://dx.doi.org/}%
\providecommand \selectlanguage [0]{\@gobble}%
\providecommand \bibinfo  [0]{\@secondoftwo}%
\providecommand \bibfield  [0]{\@secondoftwo}%
\providecommand \translation [1]{[#1]}%
\providecommand \BibitemOpen [0]{}%
\providecommand \bibitemStop [0]{}%
\providecommand \bibitemNoStop [0]{.\EOS\space}%
\providecommand \EOS [0]{\spacefactor3000\relax}%
\providecommand \BibitemShut  [1]{\csname bibitem#1\endcsname}%
\let\auto@bib@innerbib\@empty
\bibitem [{\citenamefont {Duncan}\ and\ \citenamefont
  {Thompson}(1992)}]{Duncan:1992hi}%
  \BibitemOpen
  \bibfield  {author} {\bibinfo {author} {\bibfnamefont {R.~C.}\ \bibnamefont
  {Duncan}}\ and\ \bibinfo {author} {\bibfnamefont {C.}~\bibnamefont
  {Thompson}},\ }\href {\doibase 10.1086/186413} {\bibfield  {journal}
  {\bibinfo  {journal} {Astrophys. J. Lett.}\ }\textbf {\bibinfo {volume}
  {392}},\ \bibinfo {pages} {L9} (\bibinfo {year} {1992})}\BibitemShut
  {NoStop}%
\bibitem [{\citenamefont {Thompson}\ and\ \citenamefont
  {Duncan}(1993)}]{Thompson:1993hn}%
  \BibitemOpen
  \bibfield  {author} {\bibinfo {author} {\bibfnamefont {C.}~\bibnamefont
  {Thompson}}\ and\ \bibinfo {author} {\bibfnamefont {R.~C.}\ \bibnamefont
  {Duncan}},\ }\href {\doibase 10.1086/172580} {\bibfield  {journal} {\bibinfo
  {journal} {Astrophys. J.}\ }\textbf {\bibinfo {volume} {408}},\ \bibinfo
  {pages} {194} (\bibinfo {year} {1993})}\BibitemShut {NoStop}%
\bibitem [{\citenamefont {Kouveliotou}\ \emph {et~al.}(1998)\citenamefont
  {Kouveliotou} \emph {et~al.}}]{Kouveliotou:1998ze}%
  \BibitemOpen
  \bibfield  {author} {\bibinfo {author} {\bibfnamefont {C.}~\bibnamefont
  {Kouveliotou}} \emph {et~al.},\ }\href {\doibase 10.1038/30410} {\bibfield
  {journal} {\bibinfo  {journal} {Nature}\ }\textbf {\bibinfo {volume} {393}},\
  \bibinfo {pages} {235} (\bibinfo {year} {1998})}\BibitemShut {NoStop}%
\bibitem [{\citenamefont {Kharzeev}\ \emph {et~al.}(2008)\citenamefont
  {Kharzeev}, \citenamefont {McLerran},\ and\ \citenamefont
  {Warringa}}]{Kharzeev:2007jp}%
  \BibitemOpen
  \bibfield  {author} {\bibinfo {author} {\bibfnamefont {D.~E.}\ \bibnamefont
  {Kharzeev}}, \bibinfo {author} {\bibfnamefont {L.~D.}\ \bibnamefont
  {McLerran}}, \ and\ \bibinfo {author} {\bibfnamefont {H.~J.}\ \bibnamefont
  {Warringa}},\ }\href {\doibase 10.1016/j.nuclphysa.2008.02.298} {\bibfield
  {journal} {\bibinfo  {journal} {Nucl. Phys. A}\ }\textbf {\bibinfo {volume}
  {803}},\ \bibinfo {pages} {227} (\bibinfo {year} {2008})},\ \Eprint
  {http://arxiv.org/abs/0711.0950} {arXiv:0711.0950 [hep-ph]} \BibitemShut
  {NoStop}%
\bibitem [{\citenamefont {Skokov}\ \emph {et~al.}(2009)\citenamefont {Skokov},
  \citenamefont {Illarionov},\ and\ \citenamefont {Toneev}}]{Skokov:2009qp}%
  \BibitemOpen
  \bibfield  {author} {\bibinfo {author} {\bibfnamefont {V.}~\bibnamefont
  {Skokov}}, \bibinfo {author} {\bibfnamefont {A.~Y.}\ \bibnamefont
  {Illarionov}}, \ and\ \bibinfo {author} {\bibfnamefont {V.}~\bibnamefont
  {Toneev}},\ }\href {\doibase 10.1142/S0217751X09047570} {\bibfield  {journal}
  {\bibinfo  {journal} {Int. J. Mod. Phys. A}\ }\textbf {\bibinfo {volume}
  {24}},\ \bibinfo {pages} {5925} (\bibinfo {year} {2009})},\ \Eprint
  {http://arxiv.org/abs/0907.1396} {arXiv:0907.1396 [nucl-th]} \BibitemShut
  {NoStop}%
\bibitem [{\citenamefont {Voronyuk}\ \emph {et~al.}(2011)\citenamefont
  {Voronyuk}, \citenamefont {Toneev}, \citenamefont {Cassing}, \citenamefont
  {Bratkovskaya}, \citenamefont {Konchakovski},\ and\ \citenamefont
  {Voloshin}}]{Voronyuk:2011jd}%
  \BibitemOpen
  \bibfield  {author} {\bibinfo {author} {\bibfnamefont {V.}~\bibnamefont
  {Voronyuk}}, \bibinfo {author} {\bibfnamefont {V.~D.}\ \bibnamefont
  {Toneev}}, \bibinfo {author} {\bibfnamefont {W.}~\bibnamefont {Cassing}},
  \bibinfo {author} {\bibfnamefont {E.~L.}\ \bibnamefont {Bratkovskaya}},
  \bibinfo {author} {\bibfnamefont {V.~P.}\ \bibnamefont {Konchakovski}}, \
  and\ \bibinfo {author} {\bibfnamefont {S.~A.}\ \bibnamefont {Voloshin}},\
  }\href {\doibase 10.1103/PhysRevC.83.054911} {\bibfield  {journal} {\bibinfo
  {journal} {Phys. Rev. C}\ }\textbf {\bibinfo {volume} {83}},\ \bibinfo
  {pages} {054911} (\bibinfo {year} {2011})},\ \Eprint
  {http://arxiv.org/abs/1103.4239} {arXiv:1103.4239 [nucl-th]} \BibitemShut
  {NoStop}%
\bibitem [{\citenamefont {Bzdak}\ and\ \citenamefont
  {Skokov}(2012)}]{Bzdak:2011yy}%
  \BibitemOpen
  \bibfield  {author} {\bibinfo {author} {\bibfnamefont {A.}~\bibnamefont
  {Bzdak}}\ and\ \bibinfo {author} {\bibfnamefont {V.}~\bibnamefont {Skokov}},\
  }\href {\doibase 10.1016/j.physletb.2012.02.065} {\bibfield  {journal}
  {\bibinfo  {journal} {Phys. Lett. B}\ }\textbf {\bibinfo {volume} {710}},\
  \bibinfo {pages} {171} (\bibinfo {year} {2012})},\ \Eprint
  {http://arxiv.org/abs/1111.1949} {arXiv:1111.1949 [hep-ph]} \BibitemShut
  {NoStop}%
\bibitem [{\citenamefont {Deng}\ and\ \citenamefont
  {Huang}(2012)}]{Deng:2012pc}%
  \BibitemOpen
  \bibfield  {author} {\bibinfo {author} {\bibfnamefont {W.-T.}\ \bibnamefont
  {Deng}}\ and\ \bibinfo {author} {\bibfnamefont {X.-G.}\ \bibnamefont
  {Huang}},\ }\href {\doibase 10.1103/PhysRevC.85.044907} {\bibfield  {journal}
  {\bibinfo  {journal} {Phys. Rev. C}\ }\textbf {\bibinfo {volume} {85}},\
  \bibinfo {pages} {044907} (\bibinfo {year} {2012})},\ \Eprint
  {http://arxiv.org/abs/1201.5108} {arXiv:1201.5108 [nucl-th]} \BibitemShut
  {NoStop}%
\bibitem [{\citenamefont {Inghirami}\ \emph {et~al.}(2016)\citenamefont
  {Inghirami}, \citenamefont {Del~Zanna}, \citenamefont {Beraudo},
  \citenamefont {Moghaddam}, \citenamefont {Becattini},\ and\ \citenamefont
  {Bleicher}}]{Inghirami:2016iru}%
  \BibitemOpen
  \bibfield  {author} {\bibinfo {author} {\bibfnamefont {G.}~\bibnamefont
  {Inghirami}}, \bibinfo {author} {\bibfnamefont {L.}~\bibnamefont
  {Del~Zanna}}, \bibinfo {author} {\bibfnamefont {A.}~\bibnamefont {Beraudo}},
  \bibinfo {author} {\bibfnamefont {M.~H.}\ \bibnamefont {Moghaddam}}, \bibinfo
  {author} {\bibfnamefont {F.}~\bibnamefont {Becattini}}, \ and\ \bibinfo
  {author} {\bibfnamefont {M.}~\bibnamefont {Bleicher}},\ }\href {\doibase
  10.1140/epjc/s10052-016-4516-8} {\bibfield  {journal} {\bibinfo  {journal}
  {Eur. Phys. J. C}\ }\textbf {\bibinfo {volume} {76}},\ \bibinfo {pages} {659}
  (\bibinfo {year} {2016})},\ \Eprint {http://arxiv.org/abs/1609.03042}
  {arXiv:1609.03042 [hep-ph]} \BibitemShut {NoStop}%
\bibitem [{\citenamefont {Roy}\ \emph {et~al.}(2017)\citenamefont {Roy},
  \citenamefont {Pu}, \citenamefont {Rezzolla},\ and\ \citenamefont
  {Rischke}}]{Roy:2017yvg}%
  \BibitemOpen
  \bibfield  {author} {\bibinfo {author} {\bibfnamefont {V.}~\bibnamefont
  {Roy}}, \bibinfo {author} {\bibfnamefont {S.}~\bibnamefont {Pu}}, \bibinfo
  {author} {\bibfnamefont {L.}~\bibnamefont {Rezzolla}}, \ and\ \bibinfo
  {author} {\bibfnamefont {D.~H.}\ \bibnamefont {Rischke}},\ }\href {\doibase
  10.1103/PhysRevC.96.054909} {\bibfield  {journal} {\bibinfo  {journal} {Phys.
  Rev. C}\ }\textbf {\bibinfo {volume} {96}},\ \bibinfo {pages} {054909}
  (\bibinfo {year} {2017})},\ \Eprint {http://arxiv.org/abs/1706.05326}
  {arXiv:1706.05326 [nucl-th]} \BibitemShut {NoStop}%
\bibitem [{\citenamefont {Vachaspati}(1991)}]{Vachaspati:1991nm}%
  \BibitemOpen
  \bibfield  {author} {\bibinfo {author} {\bibfnamefont {T.}~\bibnamefont
  {Vachaspati}},\ }\href {\doibase 10.1016/0370-2693(91)90051-Q} {\bibfield
  {journal} {\bibinfo  {journal} {Phys. Lett. B}\ }\textbf {\bibinfo {volume}
  {265}},\ \bibinfo {pages} {258} (\bibinfo {year} {1991})}\BibitemShut
  {NoStop}%
\bibitem [{\citenamefont {Enqvist}\ and\ \citenamefont
  {Olesen}(1993)}]{Enqvist:1993np}%
  \BibitemOpen
  \bibfield  {author} {\bibinfo {author} {\bibfnamefont {K.}~\bibnamefont
  {Enqvist}}\ and\ \bibinfo {author} {\bibfnamefont {P.}~\bibnamefont
  {Olesen}},\ }\href {\doibase 10.1016/0370-2693(93)90799-N} {\bibfield
  {journal} {\bibinfo  {journal} {Phys. Lett. B}\ }\textbf {\bibinfo {volume}
  {319}},\ \bibinfo {pages} {178} (\bibinfo {year} {1993})},\ \Eprint
  {http://arxiv.org/abs/hep-ph/9308270} {arXiv:hep-ph/9308270} \BibitemShut
  {NoStop}%
\bibitem [{\citenamefont {Grasso}\ and\ \citenamefont
  {Rubinstein}(2001)}]{Grasso:2000wj}%
  \BibitemOpen
  \bibfield  {author} {\bibinfo {author} {\bibfnamefont {D.}~\bibnamefont
  {Grasso}}\ and\ \bibinfo {author} {\bibfnamefont {H.~R.}\ \bibnamefont
  {Rubinstein}},\ }\href {\doibase 10.1016/S0370-1573(00)00110-1} {\bibfield
  {journal} {\bibinfo  {journal} {Phys. Rept.}\ }\textbf {\bibinfo {volume}
  {348}},\ \bibinfo {pages} {163} (\bibinfo {year} {2001})},\ \Eprint
  {http://arxiv.org/abs/astro-ph/0009061} {arXiv:astro-ph/0009061} \BibitemShut
  {NoStop}%
\bibitem [{\citenamefont {Aarts}(2016)}]{Aarts:2015tyj}%
  \BibitemOpen
  \bibfield  {author} {\bibinfo {author} {\bibfnamefont {G.}~\bibnamefont
  {Aarts}},\ }\href {\doibase 10.1088/1742-6596/706/2/022004} {\bibfield
  {journal} {\bibinfo  {journal} {J. Phys. Conf. Ser.}\ }\textbf {\bibinfo
  {volume} {706}},\ \bibinfo {pages} {022004} (\bibinfo {year} {2016})},\
  \Eprint {http://arxiv.org/abs/1512.05145} {arXiv:1512.05145 [hep-lat]}
  \BibitemShut {NoStop}%
\bibitem [{\citenamefont {D'Elia}\ \emph {et~al.}(2010)\citenamefont {D'Elia},
  \citenamefont {Mukherjee},\ and\ \citenamefont {Sanfilippo}}]{DElia:2010abb}%
  \BibitemOpen
  \bibfield  {author} {\bibinfo {author} {\bibfnamefont {M.}~\bibnamefont
  {D'Elia}}, \bibinfo {author} {\bibfnamefont {S.}~\bibnamefont {Mukherjee}}, \
  and\ \bibinfo {author} {\bibfnamefont {F.}~\bibnamefont {Sanfilippo}},\
  }\href {\doibase 10.1103/PhysRevD.82.051501} {\bibfield  {journal} {\bibinfo
  {journal} {Phys. Rev. D}\ }\textbf {\bibinfo {volume} {82}},\ \bibinfo
  {pages} {051501(R)} (\bibinfo {year} {2010})},\ \Eprint
  {http://arxiv.org/abs/1005.5365} {arXiv:1005.5365 [hep-lat]} \BibitemShut
  {NoStop}%
\bibitem [{\citenamefont {Bali}\ \emph
  {et~al.}(2012{\natexlab{a}})\citenamefont {Bali}, \citenamefont {Bruckmann},
  \citenamefont {Endr\"odi}, \citenamefont {Fodor}, \citenamefont {Katz},
  \citenamefont {Krieg}, \citenamefont {Schafer},\ and\ \citenamefont
  {Szabo}}]{Bali:2011qj}%
  \BibitemOpen
  \bibfield  {author} {\bibinfo {author} {\bibfnamefont {G.~S.}\ \bibnamefont
  {Bali}}, \bibinfo {author} {\bibfnamefont {F.}~\bibnamefont {Bruckmann}},
  \bibinfo {author} {\bibfnamefont {G.}~\bibnamefont {Endr\"odi}}, \bibinfo
  {author} {\bibfnamefont {Z.}~\bibnamefont {Fodor}}, \bibinfo {author}
  {\bibfnamefont {S.~D.}\ \bibnamefont {Katz}}, \bibinfo {author}
  {\bibfnamefont {S.}~\bibnamefont {Krieg}}, \bibinfo {author} {\bibfnamefont
  {A.}~\bibnamefont {Schafer}}, \ and\ \bibinfo {author} {\bibfnamefont
  {K.~K.}\ \bibnamefont {Szabo}},\ }\href {\doibase 10.1007/JHEP02(2012)044}
  {\bibfield  {journal} {\bibinfo  {journal} {JHEP}\ }\textbf {\bibinfo
  {volume} {02}},\ \bibinfo {pages} {044} (\bibinfo {year}
  {2012}{\natexlab{a}})},\ \Eprint {http://arxiv.org/abs/1111.4956}
  {arXiv:1111.4956 [hep-lat]} \BibitemShut {NoStop}%
\bibitem [{\citenamefont {Ilgenfritz}\ \emph {et~al.}(2012)\citenamefont
  {Ilgenfritz}, \citenamefont {Kalinowski}, \citenamefont {Muller-Preussker},
  \citenamefont {Petersson},\ and\ \citenamefont
  {Schreiber}}]{Ilgenfritz:2012fw}%
  \BibitemOpen
  \bibfield  {author} {\bibinfo {author} {\bibfnamefont {E.~M.}\ \bibnamefont
  {Ilgenfritz}}, \bibinfo {author} {\bibfnamefont {M.}~\bibnamefont
  {Kalinowski}}, \bibinfo {author} {\bibfnamefont {M.}~\bibnamefont
  {Muller-Preussker}}, \bibinfo {author} {\bibfnamefont {B.}~\bibnamefont
  {Petersson}}, \ and\ \bibinfo {author} {\bibfnamefont {A.}~\bibnamefont
  {Schreiber}},\ }\href {\doibase 10.1103/PhysRevD.85.114504} {\bibfield
  {journal} {\bibinfo  {journal} {Phys. Rev. D}\ }\textbf {\bibinfo {volume}
  {85}},\ \bibinfo {pages} {114504} (\bibinfo {year} {2012})},\ \Eprint
  {http://arxiv.org/abs/1203.3360} {arXiv:1203.3360 [hep-lat]} \BibitemShut
  {NoStop}%
\bibitem [{\citenamefont {Bali}\ \emph
  {et~al.}(2012{\natexlab{b}})\citenamefont {Bali}, \citenamefont {Bruckmann},
  \citenamefont {Endrodi}, \citenamefont {Fodor}, \citenamefont {Katz},\ and\
  \citenamefont {Schafer}}]{Bali:2012zg}%
  \BibitemOpen
  \bibfield  {author} {\bibinfo {author} {\bibfnamefont {G.~S.}\ \bibnamefont
  {Bali}}, \bibinfo {author} {\bibfnamefont {F.}~\bibnamefont {Bruckmann}},
  \bibinfo {author} {\bibfnamefont {G.}~\bibnamefont {Endrodi}}, \bibinfo
  {author} {\bibfnamefont {Z.}~\bibnamefont {Fodor}}, \bibinfo {author}
  {\bibfnamefont {S.~D.}\ \bibnamefont {Katz}}, \ and\ \bibinfo {author}
  {\bibfnamefont {A.}~\bibnamefont {Schafer}},\ }\href {\doibase
  10.1103/PhysRevD.86.071502} {\bibfield  {journal} {\bibinfo  {journal} {Phys.
  Rev. D}\ }\textbf {\bibinfo {volume} {86}},\ \bibinfo {pages} {071502(R)}
  (\bibinfo {year} {2012}{\natexlab{b}})},\ \Eprint
  {http://arxiv.org/abs/1206.4205} {arXiv:1206.4205 [hep-lat]} \BibitemShut
  {NoStop}%
\bibitem [{\citenamefont {Bornyakov}\ \emph {et~al.}(2014)\citenamefont
  {Bornyakov}, \citenamefont {Buividovich}, \citenamefont {Cundy},
  \citenamefont {Kochetkov},\ and\ \citenamefont
  {Sch\"afer}}]{Bornyakov:2013eya}%
  \BibitemOpen
  \bibfield  {author} {\bibinfo {author} {\bibfnamefont {V.~G.}\ \bibnamefont
  {Bornyakov}}, \bibinfo {author} {\bibfnamefont {P.~V.}\ \bibnamefont
  {Buividovich}}, \bibinfo {author} {\bibfnamefont {N.}~\bibnamefont {Cundy}},
  \bibinfo {author} {\bibfnamefont {O.~A.}\ \bibnamefont {Kochetkov}}, \ and\
  \bibinfo {author} {\bibfnamefont {A.}~\bibnamefont {Sch\"afer}},\ }\href
  {\doibase 10.1103/PhysRevD.90.034501} {\bibfield  {journal} {\bibinfo
  {journal} {Phys. Rev. D}\ }\textbf {\bibinfo {volume} {90}},\ \bibinfo
  {pages} {034501} (\bibinfo {year} {2014})},\ \Eprint
  {http://arxiv.org/abs/1312.5628} {arXiv:1312.5628 [hep-lat]} \BibitemShut
  {NoStop}%
\bibitem [{\citenamefont {Bali}\ \emph {et~al.}(2013)\citenamefont {Bali},
  \citenamefont {Bruckmann}, \citenamefont {Endrodi}, \citenamefont {Gruber},\
  and\ \citenamefont {Schaefer}}]{Bali:2013esa}%
  \BibitemOpen
  \bibfield  {author} {\bibinfo {author} {\bibfnamefont {G.~S.}\ \bibnamefont
  {Bali}}, \bibinfo {author} {\bibfnamefont {F.}~\bibnamefont {Bruckmann}},
  \bibinfo {author} {\bibfnamefont {G.}~\bibnamefont {Endrodi}}, \bibinfo
  {author} {\bibfnamefont {F.}~\bibnamefont {Gruber}}, \ and\ \bibinfo {author}
  {\bibfnamefont {A.}~\bibnamefont {Schaefer}},\ }\href {\doibase
  10.1007/JHEP04(2013)130} {\bibfield  {journal} {\bibinfo  {journal} {JHEP}\
  }\textbf {\bibinfo {volume} {04}},\ \bibinfo {pages} {130} (\bibinfo {year}
  {2013})},\ \Eprint {http://arxiv.org/abs/1303.1328} {arXiv:1303.1328
  [hep-lat]} \BibitemShut {NoStop}%
\bibitem [{\citenamefont {Bruckmann}\ \emph {et~al.}(2013)\citenamefont
  {Bruckmann}, \citenamefont {Endrodi},\ and\ \citenamefont
  {Kovacs}}]{Bruckmann:2013oba}%
  \BibitemOpen
  \bibfield  {author} {\bibinfo {author} {\bibfnamefont {F.}~\bibnamefont
  {Bruckmann}}, \bibinfo {author} {\bibfnamefont {G.}~\bibnamefont {Endrodi}},
  \ and\ \bibinfo {author} {\bibfnamefont {T.~G.}\ \bibnamefont {Kovacs}},\
  }\href {\doibase 10.1007/JHEP04(2013)112} {\bibfield  {journal} {\bibinfo
  {journal} {JHEP}\ }\textbf {\bibinfo {volume} {04}},\ \bibinfo {pages} {112}
  (\bibinfo {year} {2013})},\ \Eprint {http://arxiv.org/abs/1303.3972}
  {arXiv:1303.3972 [hep-lat]} \BibitemShut {NoStop}%
\bibitem [{\citenamefont {Bali}\ \emph {et~al.}(2014)\citenamefont {Bali},
  \citenamefont {Bruckmann}, \citenamefont {Endr\"odi}, \citenamefont {Katz},\
  and\ \citenamefont {Sch\"afer}}]{Bali:2014kia}%
  \BibitemOpen
  \bibfield  {author} {\bibinfo {author} {\bibfnamefont {G.~S.}\ \bibnamefont
  {Bali}}, \bibinfo {author} {\bibfnamefont {F.}~\bibnamefont {Bruckmann}},
  \bibinfo {author} {\bibfnamefont {G.}~\bibnamefont {Endr\"odi}}, \bibinfo
  {author} {\bibfnamefont {S.~D.}\ \bibnamefont {Katz}}, \ and\ \bibinfo
  {author} {\bibfnamefont {A.}~\bibnamefont {Sch\"afer}},\ }\href {\doibase
  10.1007/JHEP08(2014)177} {\bibfield  {journal} {\bibinfo  {journal} {JHEP}\
  }\textbf {\bibinfo {volume} {08}},\ \bibinfo {pages} {177} (\bibinfo {year}
  {2014})},\ \Eprint {http://arxiv.org/abs/1406.0269} {arXiv:1406.0269
  [hep-lat]} \BibitemShut {NoStop}%
\bibitem [{\citenamefont {Endr\"odi}(2015)}]{Endrodi:2015oba}%
  \BibitemOpen
  \bibfield  {author} {\bibinfo {author} {\bibfnamefont {G.}~\bibnamefont
  {Endr\"odi}},\ }\href {\doibase 10.1007/JHEP07(2015)173} {\bibfield
  {journal} {\bibinfo  {journal} {JHEP}\ }\textbf {\bibinfo {volume} {07}},\
  \bibinfo {pages} {173} (\bibinfo {year} {2015})},\ \Eprint
  {http://arxiv.org/abs/1504.08280} {arXiv:1504.08280 [hep-lat]} \BibitemShut
  {NoStop}%
\bibitem [{\citenamefont {D'Elia}\ \emph {et~al.}(2018)\citenamefont {D'Elia},
  \citenamefont {Manigrasso}, \citenamefont {Negro},\ and\ \citenamefont
  {Sanfilippo}}]{DElia:2018xwo}%
  \BibitemOpen
  \bibfield  {author} {\bibinfo {author} {\bibfnamefont {M.}~\bibnamefont
  {D'Elia}}, \bibinfo {author} {\bibfnamefont {F.}~\bibnamefont {Manigrasso}},
  \bibinfo {author} {\bibfnamefont {F.}~\bibnamefont {Negro}}, \ and\ \bibinfo
  {author} {\bibfnamefont {F.}~\bibnamefont {Sanfilippo}},\ }\href {\doibase
  10.1103/PhysRevD.98.054509} {\bibfield  {journal} {\bibinfo  {journal} {Phys.
  Rev. D}\ }\textbf {\bibinfo {volume} {98}},\ \bibinfo {pages} {054509}
  (\bibinfo {year} {2018})},\ \Eprint {http://arxiv.org/abs/1808.07008}
  {arXiv:1808.07008 [hep-lat]} \BibitemShut {NoStop}%
\bibitem [{\citenamefont {Blaizot}\ \emph {et~al.}(2013)\citenamefont
  {Blaizot}, \citenamefont {Fraga},\ and\ \citenamefont
  {Palhares}}]{Blaizot:2012sd}%
  \BibitemOpen
  \bibfield  {author} {\bibinfo {author} {\bibfnamefont {J.-P.}\ \bibnamefont
  {Blaizot}}, \bibinfo {author} {\bibfnamefont {E.~S.}\ \bibnamefont {Fraga}},
  \ and\ \bibinfo {author} {\bibfnamefont {L.~F.}\ \bibnamefont {Palhares}},\
  }\href {\doibase 10.1016/j.physletb.2013.04.004} {\bibfield  {journal}
  {\bibinfo  {journal} {Phys. Lett. B}\ }\textbf {\bibinfo {volume} {722}},\
  \bibinfo {pages} {167} (\bibinfo {year} {2013})},\ \Eprint
  {http://arxiv.org/abs/1211.6412} {arXiv:1211.6412 [hep-ph]} \BibitemShut
  {NoStop}%
\bibitem [{\citenamefont {Ayala}\ \emph {et~al.}(2015)\citenamefont {Ayala},
  \citenamefont {Cobos-Mart\'\i{}nez}, \citenamefont {Loewe}, \citenamefont
  {Tejeda-Yeomans},\ and\ \citenamefont {Zamora}}]{Ayala:2014uua}%
  \BibitemOpen
  \bibfield  {author} {\bibinfo {author} {\bibfnamefont {A.}~\bibnamefont
  {Ayala}}, \bibinfo {author} {\bibfnamefont {J.~J.}\ \bibnamefont
  {Cobos-Mart\'\i{}nez}}, \bibinfo {author} {\bibfnamefont {M.}~\bibnamefont
  {Loewe}}, \bibinfo {author} {\bibfnamefont {M.~E.}\ \bibnamefont
  {Tejeda-Yeomans}}, \ and\ \bibinfo {author} {\bibfnamefont {R.}~\bibnamefont
  {Zamora}},\ }\href {\doibase 10.1103/PhysRevD.91.016007} {\bibfield
  {journal} {\bibinfo  {journal} {Phys. Rev. D}\ }\textbf {\bibinfo {volume}
  {91}},\ \bibinfo {pages} {016007} (\bibinfo {year} {2015})},\ \Eprint
  {http://arxiv.org/abs/1410.6388} {arXiv:1410.6388 [hep-ph]} \BibitemShut
  {NoStop}%
\bibitem [{\citenamefont {Ayala}\ \emph {et~al.}(2016)\citenamefont {Ayala},
  \citenamefont {Dominguez}, \citenamefont {Hernandez}, \citenamefont {Loewe},\
  and\ \citenamefont {Zamora}}]{Ayala:2015bgv}%
  \BibitemOpen
  \bibfield  {author} {\bibinfo {author} {\bibfnamefont {A.}~\bibnamefont
  {Ayala}}, \bibinfo {author} {\bibfnamefont {C.~A.}\ \bibnamefont
  {Dominguez}}, \bibinfo {author} {\bibfnamefont {L.~A.}\ \bibnamefont
  {Hernandez}}, \bibinfo {author} {\bibfnamefont {M.}~\bibnamefont {Loewe}}, \
  and\ \bibinfo {author} {\bibfnamefont {R.}~\bibnamefont {Zamora}},\ }\href
  {\doibase 10.1016/j.physletb.2016.05.058} {\bibfield  {journal} {\bibinfo
  {journal} {Phys. Lett. B}\ }\textbf {\bibinfo {volume} {759}},\ \bibinfo
  {pages} {99} (\bibinfo {year} {2016})},\ \Eprint
  {http://arxiv.org/abs/1510.09134} {arXiv:1510.09134 [hep-ph]} \BibitemShut
  {NoStop}%
\bibitem [{\citenamefont {Rath}\ and\ \citenamefont
  {Patra}(2017)}]{Rath:2017fdv}%
  \BibitemOpen
  \bibfield  {author} {\bibinfo {author} {\bibfnamefont {S.}~\bibnamefont
  {Rath}}\ and\ \bibinfo {author} {\bibfnamefont {B.~K.}\ \bibnamefont
  {Patra}},\ }\href {\doibase 10.1007/JHEP12(2017)098} {\bibfield  {journal}
  {\bibinfo  {journal} {JHEP}\ }\textbf {\bibinfo {volume} {12}},\ \bibinfo
  {pages} {098} (\bibinfo {year} {2017})},\ \Eprint
  {http://arxiv.org/abs/1707.02890} {arXiv:1707.02890 [hep-th]} \BibitemShut
  {NoStop}%
\bibitem [{\citenamefont {Haque}(2017)}]{Haque:2017nxq}%
  \BibitemOpen
  \bibfield  {author} {\bibinfo {author} {\bibfnamefont {N.}~\bibnamefont
  {Haque}},\ }\href {\doibase 10.1103/PhysRevD.96.014019} {\bibfield  {journal}
  {\bibinfo  {journal} {Phys. Rev. D}\ }\textbf {\bibinfo {volume} {96}},\
  \bibinfo {pages} {014019} (\bibinfo {year} {2017})},\ \Eprint
  {http://arxiv.org/abs/1704.05833} {arXiv:1704.05833 [hep-ph]} \BibitemShut
  {NoStop}%
\bibitem [{\citenamefont {Karmakar}\ \emph {et~al.}(2019)\citenamefont
  {Karmakar}, \citenamefont {Ghosh}, \citenamefont {Bandyopadhyay},
  \citenamefont {Haque},\ and\ \citenamefont {Mustafa}}]{Karmakar:2019tdp}%
  \BibitemOpen
  \bibfield  {author} {\bibinfo {author} {\bibfnamefont {B.}~\bibnamefont
  {Karmakar}}, \bibinfo {author} {\bibfnamefont {R.}~\bibnamefont {Ghosh}},
  \bibinfo {author} {\bibfnamefont {A.}~\bibnamefont {Bandyopadhyay}}, \bibinfo
  {author} {\bibfnamefont {N.}~\bibnamefont {Haque}}, \ and\ \bibinfo {author}
  {\bibfnamefont {M.~G.}\ \bibnamefont {Mustafa}},\ }\href {\doibase
  10.1103/PhysRevD.99.094002} {\bibfield  {journal} {\bibinfo  {journal} {Phys.
  Rev. D}\ }\textbf {\bibinfo {volume} {99}},\ \bibinfo {pages} {094002}
  (\bibinfo {year} {2019})},\ \Eprint {http://arxiv.org/abs/1902.02607}
  {arXiv:1902.02607 [hep-ph]} \BibitemShut {NoStop}%
\bibitem [{\citenamefont {Bandyopadhyay}\ \emph {et~al.}(2019)\citenamefont
  {Bandyopadhyay}, \citenamefont {Karmakar}, \citenamefont {Haque},\ and\
  \citenamefont {Mustafa}}]{Bandyopadhyay:2017cle}%
  \BibitemOpen
  \bibfield  {author} {\bibinfo {author} {\bibfnamefont {A.}~\bibnamefont
  {Bandyopadhyay}}, \bibinfo {author} {\bibfnamefont {B.}~\bibnamefont
  {Karmakar}}, \bibinfo {author} {\bibfnamefont {N.}~\bibnamefont {Haque}}, \
  and\ \bibinfo {author} {\bibfnamefont {M.~G.}\ \bibnamefont {Mustafa}},\
  }\href {\doibase 10.1103/PhysRevD.100.034031} {\bibfield  {journal} {\bibinfo
   {journal} {Phys. Rev. D}\ }\textbf {\bibinfo {volume} {100}},\ \bibinfo
  {pages} {034031} (\bibinfo {year} {2019})},\ \Eprint
  {http://arxiv.org/abs/1702.02875} {arXiv:1702.02875 [hep-ph]} \BibitemShut
  {NoStop}%
\bibitem [{\citenamefont {Karmakar}\ \emph {et~al.}(2020)\citenamefont
  {Karmakar}, \citenamefont {Haque},\ and\ \citenamefont
  {Mustafa}}]{Karmakar:2020mnj}%
  \BibitemOpen
  \bibfield  {author} {\bibinfo {author} {\bibfnamefont {B.}~\bibnamefont
  {Karmakar}}, \bibinfo {author} {\bibfnamefont {N.}~\bibnamefont {Haque}}, \
  and\ \bibinfo {author} {\bibfnamefont {M.~G.}\ \bibnamefont {Mustafa}},\
  }\href {\doibase 10.1103/PhysRevD.102.054004} {\bibfield  {journal} {\bibinfo
   {journal} {Phys. Rev. D}\ }\textbf {\bibinfo {volume} {102}},\ \bibinfo
  {pages} {054004} (\bibinfo {year} {2020})},\ \Eprint
  {http://arxiv.org/abs/2003.11247} {arXiv:2003.11247 [hep-ph]} \BibitemShut
  {NoStop}%
\bibitem [{\citenamefont {Fraga}\ \emph {et~al.}(2013)\citenamefont {Fraga},
  \citenamefont {Noronha},\ and\ \citenamefont {Palhares}}]{Fraga:2012ev}%
  \BibitemOpen
  \bibfield  {author} {\bibinfo {author} {\bibfnamefont {E.~S.}\ \bibnamefont
  {Fraga}}, \bibinfo {author} {\bibfnamefont {J.}~\bibnamefont {Noronha}}, \
  and\ \bibinfo {author} {\bibfnamefont {L.~F.}\ \bibnamefont {Palhares}},\
  }\href {\doibase 10.1103/PhysRevD.87.114014} {\bibfield  {journal} {\bibinfo
  {journal} {Phys. Rev. D}\ }\textbf {\bibinfo {volume} {87}},\ \bibinfo
  {pages} {114014} (\bibinfo {year} {2013})},\ \Eprint
  {http://arxiv.org/abs/1207.7094} {arXiv:1207.7094 [hep-ph]} \BibitemShut
  {NoStop}%
\bibitem [{\citenamefont {Colucci}\ \emph {et~al.}(2014)\citenamefont
  {Colucci}, \citenamefont {Fraga},\ and\ \citenamefont
  {Sedrakian}}]{Colucci:2013zoa}%
  \BibitemOpen
  \bibfield  {author} {\bibinfo {author} {\bibfnamefont {G.}~\bibnamefont
  {Colucci}}, \bibinfo {author} {\bibfnamefont {E.~S.}\ \bibnamefont {Fraga}},
  \ and\ \bibinfo {author} {\bibfnamefont {A.}~\bibnamefont {Sedrakian}},\
  }\href {\doibase 10.1016/j.physletb.2013.11.028} {\bibfield  {journal}
  {\bibinfo  {journal} {Phys. Lett. B}\ }\textbf {\bibinfo {volume} {728}},\
  \bibinfo {pages} {19} (\bibinfo {year} {2014})},\ \Eprint
  {http://arxiv.org/abs/1310.3742} {arXiv:1310.3742 [nucl-th]} \BibitemShut
  {NoStop}%
\bibitem [{\citenamefont {Hofmann}(2020{\natexlab{a}})}]{Hofmann:2020dvz}%
  \BibitemOpen
  \bibfield  {author} {\bibinfo {author} {\bibfnamefont {C.~P.}\ \bibnamefont
  {Hofmann}},\ }\href {\doibase 10.1103/PhysRevD.101.114031} {\bibfield
  {journal} {\bibinfo  {journal} {Phys. Rev. D}\ }\textbf {\bibinfo {volume}
  {101}},\ \bibinfo {pages} {114031} (\bibinfo {year} {2020}{\natexlab{a}})},\
  \Eprint {http://arxiv.org/abs/2004.01247} {arXiv:2004.01247 [hep-ph]}
  \BibitemShut {NoStop}%
\bibitem [{\citenamefont {Hofmann}(2020{\natexlab{b}})}]{Hofmann:2020ism}%
  \BibitemOpen
  \bibfield  {author} {\bibinfo {author} {\bibfnamefont {C.~P.}\ \bibnamefont
  {Hofmann}},\ }\href {\doibase 10.1103/PhysRevD.102.094010} {\bibfield
  {journal} {\bibinfo  {journal} {Phys. Rev. D}\ }\textbf {\bibinfo {volume}
  {102}},\ \bibinfo {pages} {094010} (\bibinfo {year} {2020}{\natexlab{b}})},\
  \Eprint {http://arxiv.org/abs/2006.07717} {arXiv:2006.07717 [hep-ph]}
  \BibitemShut {NoStop}%
\bibitem [{\citenamefont {Fraga}(2013)}]{Fraga:2012rr}%
  \BibitemOpen
  \bibfield  {author} {\bibinfo {author} {\bibfnamefont {E.~S.}\ \bibnamefont
  {Fraga}},\ }\href {\doibase 10.1007/978-3-642-37305-3_5} {\bibfield
  {journal} {\bibinfo  {journal} {Lect. Notes Phys.}\ }\textbf {\bibinfo
  {volume} {871}},\ \bibinfo {pages} {121} (\bibinfo {year} {2013})},\ \Eprint
  {http://arxiv.org/abs/1208.0917} {arXiv:1208.0917 [hep-ph]} \BibitemShut
  {NoStop}%
\bibitem [{\citenamefont {Kharzeev}\ \emph {et~al.}(2013)\citenamefont
  {Kharzeev}, \citenamefont {Landsteiner}, \citenamefont {Schmitt},\ and\
  \citenamefont {Yee}}]{Kharzeev:2013jha}%
  \BibitemOpen
  \bibinfo {editor} {\bibfnamefont {D.}~\bibnamefont {Kharzeev}}, \bibinfo
  {editor} {\bibfnamefont {K.}~\bibnamefont {Landsteiner}}, \bibinfo {editor}
  {\bibfnamefont {A.}~\bibnamefont {Schmitt}}, \ and\ \bibinfo {editor}
  {\bibfnamefont {H.-U.}\ \bibnamefont {Yee}},\ eds.,\ \href {\doibase
  10.1007/978-3-642-37305-3} {\emph {\bibinfo {title} {{Strongly Interacting
  Matter in Magnetic Fields}}}},\ Vol.\ \bibinfo {volume} {871}\ (\bibinfo
  {year} {2013})\BibitemShut {NoStop}%
\bibitem [{\citenamefont {Andersen}\ \emph {et~al.}(2016)\citenamefont
  {Andersen}, \citenamefont {Naylor},\ and\ \citenamefont
  {Tranberg}}]{Andersen:2014xxa}%
  \BibitemOpen
  \bibfield  {author} {\bibinfo {author} {\bibfnamefont {J.~O.}\ \bibnamefont
  {Andersen}}, \bibinfo {author} {\bibfnamefont {W.~R.}\ \bibnamefont
  {Naylor}}, \ and\ \bibinfo {author} {\bibfnamefont {A.}~\bibnamefont
  {Tranberg}},\ }\href {\doibase 10.1103/RevModPhys.88.025001} {\bibfield
  {journal} {\bibinfo  {journal} {Rev. Mod. Phys.}\ }\textbf {\bibinfo {volume}
  {88}},\ \bibinfo {pages} {025001} (\bibinfo {year} {2016})},\ \Eprint
  {http://arxiv.org/abs/1411.7176} {arXiv:1411.7176 [hep-ph]} \BibitemShut
  {NoStop}%
\bibitem [{\citenamefont {Miransky}\ and\ \citenamefont
  {Shovkovy}(2015)}]{Miransky:2015ava}%
  \BibitemOpen
  \bibfield  {author} {\bibinfo {author} {\bibfnamefont {V.~A.}\ \bibnamefont
  {Miransky}}\ and\ \bibinfo {author} {\bibfnamefont {I.~A.}\ \bibnamefont
  {Shovkovy}},\ }\href {\doibase 10.1016/j.physrep.2015.02.003} {\bibfield
  {journal} {\bibinfo  {journal} {Phys. Rept.}\ }\textbf {\bibinfo {volume}
  {576}},\ \bibinfo {pages} {1} (\bibinfo {year} {2015})},\ \Eprint
  {http://arxiv.org/abs/1503.00732} {arXiv:1503.00732 [hep-ph]} \BibitemShut
  {NoStop}%
\bibitem [{\citenamefont {D'Elia}\ \emph {et~al.}(2021)\citenamefont {D'Elia},
  \citenamefont {Maio}, \citenamefont {Sanfilippo},\ and\ \citenamefont
  {Stanzione}}]{DElia:2021tfb}%
  \BibitemOpen
  \bibfield  {author} {\bibinfo {author} {\bibfnamefont {M.}~\bibnamefont
  {D'Elia}}, \bibinfo {author} {\bibfnamefont {L.}~\bibnamefont {Maio}},
  \bibinfo {author} {\bibfnamefont {F.}~\bibnamefont {Sanfilippo}}, \ and\
  \bibinfo {author} {\bibfnamefont {A.}~\bibnamefont {Stanzione}},\ }\href
  {\doibase 10.1103/PhysRevD.104.114512} {\bibfield  {journal} {\bibinfo
  {journal} {Phys. Rev. D}\ }\textbf {\bibinfo {volume} {104}},\ \bibinfo
  {pages} {114512} (\bibinfo {year} {2021})},\ \Eprint
  {http://arxiv.org/abs/2109.07456} {arXiv:2109.07456 [hep-lat]} \BibitemShut
  {NoStop}%
\bibitem [{\citenamefont {D'Elia}\ \emph {et~al.}(2022)\citenamefont {D'Elia},
  \citenamefont {Maio}, \citenamefont {Sanfilippo},\ and\ \citenamefont
  {Stanzione}}]{DElia:2021yvk}%
  \BibitemOpen
  \bibfield  {author} {\bibinfo {author} {\bibfnamefont {M.}~\bibnamefont
  {D'Elia}}, \bibinfo {author} {\bibfnamefont {L.}~\bibnamefont {Maio}},
  \bibinfo {author} {\bibfnamefont {F.}~\bibnamefont {Sanfilippo}}, \ and\
  \bibinfo {author} {\bibfnamefont {A.}~\bibnamefont {Stanzione}},\ }\href
  {\doibase 10.1103/PhysRevD.105.034511} {\bibfield  {journal} {\bibinfo
  {journal} {Phys. Rev. D}\ }\textbf {\bibinfo {volume} {105}},\ \bibinfo
  {pages} {034511} (\bibinfo {year} {2022})},\ \Eprint
  {http://arxiv.org/abs/2111.11237} {arXiv:2111.11237 [hep-lat]} \BibitemShut
  {NoStop}%
\bibitem [{\citenamefont {Fraga}\ and\ \citenamefont
  {Mizher}(2008)}]{Fraga:2008qn}%
  \BibitemOpen
  \bibfield  {author} {\bibinfo {author} {\bibfnamefont {E.~S.}\ \bibnamefont
  {Fraga}}\ and\ \bibinfo {author} {\bibfnamefont {A.~J.}\ \bibnamefont
  {Mizher}},\ }\href {\doibase 10.1103/PhysRevD.78.025016} {\bibfield
  {journal} {\bibinfo  {journal} {Phys. Rev. D}\ }\textbf {\bibinfo {volume}
  {78}},\ \bibinfo {pages} {025016} (\bibinfo {year} {2008})},\ \Eprint
  {http://arxiv.org/abs/0804.1452} {arXiv:0804.1452 [hep-ph]} \BibitemShut
  {NoStop}%
\bibitem [{\citenamefont {Mizher}\ \emph {et~al.}(2010)\citenamefont {Mizher},
  \citenamefont {Chernodub},\ and\ \citenamefont {Fraga}}]{Mizher:2010zb}%
  \BibitemOpen
  \bibfield  {author} {\bibinfo {author} {\bibfnamefont {A.~J.}\ \bibnamefont
  {Mizher}}, \bibinfo {author} {\bibfnamefont {M.~N.}\ \bibnamefont
  {Chernodub}}, \ and\ \bibinfo {author} {\bibfnamefont {E.~S.}\ \bibnamefont
  {Fraga}},\ }\href {\doibase 10.1103/PhysRevD.82.105016} {\bibfield  {journal}
  {\bibinfo  {journal} {Phys. Rev. D}\ }\textbf {\bibinfo {volume} {82}},\
  \bibinfo {pages} {105016} (\bibinfo {year} {2010})},\ \Eprint
  {http://arxiv.org/abs/1004.2712} {arXiv:1004.2712 [hep-ph]} \BibitemShut
  {NoStop}%
\bibitem [{\citenamefont {Fraga}\ and\ \citenamefont
  {Palhares}(2012)}]{Fraga:2012fs}%
  \BibitemOpen
  \bibfield  {author} {\bibinfo {author} {\bibfnamefont {E.~S.}\ \bibnamefont
  {Fraga}}\ and\ \bibinfo {author} {\bibfnamefont {L.~F.}\ \bibnamefont
  {Palhares}},\ }\href {\doibase 10.1103/PhysRevD.86.016008} {\bibfield
  {journal} {\bibinfo  {journal} {Phys. Rev. D}\ }\textbf {\bibinfo {volume}
  {86}},\ \bibinfo {pages} {016008} (\bibinfo {year} {2012})},\ \Eprint
  {http://arxiv.org/abs/1201.5881} {arXiv:1201.5881 [hep-ph]} \BibitemShut
  {NoStop}%
\bibitem [{\citenamefont {Endr\"odi}(2013)}]{Endrodi:2013cs}%
  \BibitemOpen
  \bibfield  {author} {\bibinfo {author} {\bibfnamefont {G.}~\bibnamefont
  {Endr\"odi}},\ }\href {\doibase 10.1007/JHEP04(2013)023} {\bibfield
  {journal} {\bibinfo  {journal} {JHEP}\ }\textbf {\bibinfo {volume} {04}},\
  \bibinfo {pages} {023} (\bibinfo {year} {2013})},\ \Eprint
  {http://arxiv.org/abs/1301.1307} {arXiv:1301.1307 [hep-ph]} \BibitemShut
  {NoStop}%
\bibitem [{\citenamefont {Haber}\ \emph {et~al.}(2014)\citenamefont {Haber},
  \citenamefont {Preis},\ and\ \citenamefont {Schmitt}}]{Haber:2014ula}%
  \BibitemOpen
  \bibfield  {author} {\bibinfo {author} {\bibfnamefont {A.}~\bibnamefont
  {Haber}}, \bibinfo {author} {\bibfnamefont {F.}~\bibnamefont {Preis}}, \ and\
  \bibinfo {author} {\bibfnamefont {A.}~\bibnamefont {Schmitt}},\ }\href
  {\doibase 10.1103/PhysRevD.90.125036} {\bibfield  {journal} {\bibinfo
  {journal} {Phys. Rev. D}\ }\textbf {\bibinfo {volume} {90}},\ \bibinfo
  {pages} {125036} (\bibinfo {year} {2014})},\ \Eprint
  {http://arxiv.org/abs/1409.0425} {arXiv:1409.0425 [nucl-th]} \BibitemShut
  {NoStop}%
\bibitem [{\citenamefont {Avancini}\ \emph {et~al.}(2021)\citenamefont
  {Avancini}, \citenamefont {Farias}, \citenamefont {Pinto}, \citenamefont
  {Restrepo},\ and\ \citenamefont {Tavares}}]{Avancini:2020xqe}%
  \BibitemOpen
  \bibfield  {author} {\bibinfo {author} {\bibfnamefont {S.~S.}\ \bibnamefont
  {Avancini}}, \bibinfo {author} {\bibfnamefont {R.~L.~S.}\ \bibnamefont
  {Farias}}, \bibinfo {author} {\bibfnamefont {M.~B.}\ \bibnamefont {Pinto}},
  \bibinfo {author} {\bibfnamefont {T.~E.}\ \bibnamefont {Restrepo}}, \ and\
  \bibinfo {author} {\bibfnamefont {W.~R.}\ \bibnamefont {Tavares}},\ }\href
  {\doibase 10.1103/PhysRevD.103.056009} {\bibfield  {journal} {\bibinfo
  {journal} {Phys. Rev. D}\ }\textbf {\bibinfo {volume} {103}},\ \bibinfo
  {pages} {056009} (\bibinfo {year} {2021})},\ \Eprint
  {http://arxiv.org/abs/2008.10720} {arXiv:2008.10720 [hep-ph]} \BibitemShut
  {NoStop}%
\bibitem [{\citenamefont {Tavares}\ \emph {et~al.}(2021)\citenamefont
  {Tavares}, \citenamefont {Farias}, \citenamefont {Avancini}, \citenamefont
  {Tim\'oteo}, \citenamefont {Pinto},\ and\ \citenamefont
  {Krein}}]{Tavares:2021fik}%
  \BibitemOpen
  \bibfield  {author} {\bibinfo {author} {\bibfnamefont {W.~R.}\ \bibnamefont
  {Tavares}}, \bibinfo {author} {\bibfnamefont {R.~L.~S.}\ \bibnamefont
  {Farias}}, \bibinfo {author} {\bibfnamefont {S.~S.}\ \bibnamefont
  {Avancini}}, \bibinfo {author} {\bibfnamefont {V.~S.}\ \bibnamefont
  {Tim\'oteo}}, \bibinfo {author} {\bibfnamefont {M.~B.}\ \bibnamefont
  {Pinto}}, \ and\ \bibinfo {author} {\bibfnamefont {G.~a.}\ \bibnamefont
  {Krein}},\ }\href {\doibase 10.1140/epja/s10050-021-00587-5} {\bibfield
  {journal} {\bibinfo  {journal} {Eur. Phys. J. A}\ }\textbf {\bibinfo {volume}
  {57}},\ \bibinfo {pages} {278} (\bibinfo {year} {2021})},\ \Eprint
  {http://arxiv.org/abs/2104.11117} {arXiv:2104.11117 [hep-ph]} \BibitemShut
  {NoStop}%
\bibitem [{\citenamefont {Farias}\ \emph {et~al.}(2022)\citenamefont {Farias},
  \citenamefont {Tavares}, \citenamefont {Nunes},\ and\ \citenamefont
  {Avancini}}]{Farias:2021fci}%
  \BibitemOpen
  \bibfield  {author} {\bibinfo {author} {\bibfnamefont {R.~L.~S.}\
  \bibnamefont {Farias}}, \bibinfo {author} {\bibfnamefont {W.~R.}\
  \bibnamefont {Tavares}}, \bibinfo {author} {\bibfnamefont {R.~M.}\
  \bibnamefont {Nunes}}, \ and\ \bibinfo {author} {\bibfnamefont {S.~S.}\
  \bibnamefont {Avancini}},\ }\href {\doibase 10.1140/epjc/s10052-022-10640-2}
  {\bibfield  {journal} {\bibinfo  {journal} {Eur. Phys. J. C}\ }\textbf
  {\bibinfo {volume} {82}},\ \bibinfo {pages} {674} (\bibinfo {year} {2022})},\
  \Eprint {http://arxiv.org/abs/2109.11112} {arXiv:2109.11112 [hep-ph]}
  \BibitemShut {NoStop}%
\bibitem [{\citenamefont {Kapusta}\ and\ \citenamefont
  {Gale}(2011)}]{Kapusta:2006pm}%
  \BibitemOpen
  \bibfield  {author} {\bibinfo {author} {\bibfnamefont {J.~I.}\ \bibnamefont
  {Kapusta}}\ and\ \bibinfo {author} {\bibfnamefont {C.}~\bibnamefont {Gale}},\
  }\href {\doibase 10.1017/CBO9780511535130} {\emph {\bibinfo {title}
  {{Finite-temperature field theory: Principles and applications}}}},\
  Cambridge Monographs on Mathematical Physics\ (\bibinfo  {publisher}
  {Cambridge University Press},\ \bibinfo {year} {2011})\BibitemShut {NoStop}%
\bibitem [{\citenamefont {Vermaseren}\ \emph {et~al.}(1997)\citenamefont
  {Vermaseren}, \citenamefont {Larin},\ and\ \citenamefont {van
  Ritbergen}}]{Vermaseren:1997fq}%
  \BibitemOpen
  \bibfield  {author} {\bibinfo {author} {\bibfnamefont {J.~A.~M.}\
  \bibnamefont {Vermaseren}}, \bibinfo {author} {\bibfnamefont {S.~A.}\
  \bibnamefont {Larin}}, \ and\ \bibinfo {author} {\bibfnamefont
  {T.}~\bibnamefont {van Ritbergen}},\ }\href {\doibase
  10.1016/S0370-2693(97)00660-6} {\bibfield  {journal} {\bibinfo  {journal}
  {Phys. Lett. B}\ }\textbf {\bibinfo {volume} {405}},\ \bibinfo {pages} {327}
  (\bibinfo {year} {1997})},\ \Eprint {http://arxiv.org/abs/hep-ph/9703284}
  {arXiv:hep-ph/9703284} \BibitemShut {NoStop}%
\bibitem [{\citenamefont {Fraga}\ and\ \citenamefont
  {Romatschke}(2005)}]{Fraga:2004gz}%
  \BibitemOpen
  \bibfield  {author} {\bibinfo {author} {\bibfnamefont {E.~S.}\ \bibnamefont
  {Fraga}}\ and\ \bibinfo {author} {\bibfnamefont {P.}~\bibnamefont
  {Romatschke}},\ }\href {\doibase 10.1103/PhysRevD.71.105014} {\bibfield
  {journal} {\bibinfo  {journal} {Phys. Rev. D}\ }\textbf {\bibinfo {volume}
  {71}},\ \bibinfo {pages} {105014} (\bibinfo {year} {2005})},\ \Eprint
  {http://arxiv.org/abs/hep-ph/0412298} {arXiv:hep-ph/0412298} \BibitemShut
  {NoStop}%
\bibitem [{\citenamefont {Bazavov}\ \emph {et~al.}(2014)\citenamefont
  {Bazavov}, \citenamefont {Brambilla}, \citenamefont {Garcia~i Tormo},
  \citenamefont {Petreczky}, \citenamefont {Soto},\ and\ \citenamefont
  {Vairo}}]{Bazavov:2014soa}%
  \BibitemOpen
  \bibfield  {author} {\bibinfo {author} {\bibfnamefont {A.}~\bibnamefont
  {Bazavov}}, \bibinfo {author} {\bibfnamefont {N.}~\bibnamefont {Brambilla}},
  \bibinfo {author} {\bibfnamefont {X.}~\bibnamefont {Garcia~i Tormo}},
  \bibinfo {author} {\bibfnamefont {P.}~\bibnamefont {Petreczky}}, \bibinfo
  {author} {\bibfnamefont {J.}~\bibnamefont {Soto}}, \ and\ \bibinfo {author}
  {\bibfnamefont {A.}~\bibnamefont {Vairo}},\ }\href {\doibase
  10.1103/PhysRevD.90.074038} {\bibfield  {journal} {\bibinfo  {journal} {Phys.
  Rev. D}\ }\textbf {\bibinfo {volume} {90}},\ \bibinfo {pages} {074038}
  (\bibinfo {year} {2014})},\ \bibinfo {note} {[Erratum: Phys.Rev.D 101,
  119902(E) (2020)]},\ \Eprint {http://arxiv.org/abs/1407.8437}
  {arXiv:1407.8437 [hep-ph]} \BibitemShut {NoStop}%
\bibitem [{\citenamefont {Chakraborty}\ \emph {et~al.}(2015)\citenamefont
  {Chakraborty}, \citenamefont {Davies}, \citenamefont {Galloway},
  \citenamefont {Knecht}, \citenamefont {Koponen}, \citenamefont {Donald},
  \citenamefont {Dowdall}, \citenamefont {Lepage},\ and\ \citenamefont
  {McNeile}}]{Chakraborty:2014aca}%
  \BibitemOpen
  \bibfield  {author} {\bibinfo {author} {\bibfnamefont {B.}~\bibnamefont
  {Chakraborty}}, \bibinfo {author} {\bibfnamefont {C.~T.~H.}\ \bibnamefont
  {Davies}}, \bibinfo {author} {\bibfnamefont {B.}~\bibnamefont {Galloway}},
  \bibinfo {author} {\bibfnamefont {P.}~\bibnamefont {Knecht}}, \bibinfo
  {author} {\bibfnamefont {J.}~\bibnamefont {Koponen}}, \bibinfo {author}
  {\bibfnamefont {G.~C.}\ \bibnamefont {Donald}}, \bibinfo {author}
  {\bibfnamefont {R.~J.}\ \bibnamefont {Dowdall}}, \bibinfo {author}
  {\bibfnamefont {G.~P.}\ \bibnamefont {Lepage}}, \ and\ \bibinfo {author}
  {\bibfnamefont {C.}~\bibnamefont {McNeile}},\ }\href {\doibase
  10.1103/PhysRevD.91.054508} {\bibfield  {journal} {\bibinfo  {journal} {Phys.
  Rev. D}\ }\textbf {\bibinfo {volume} {91}},\ \bibinfo {pages} {054508}
  (\bibinfo {year} {2015})},\ \Eprint {http://arxiv.org/abs/1408.4169}
  {arXiv:1408.4169 [hep-lat]} \BibitemShut {NoStop}%
\bibitem [{\citenamefont {Ayala}\ \emph {et~al.}(2018)\citenamefont {Ayala},
  \citenamefont {Dominguez}, \citenamefont {Hernandez-Ortiz}, \citenamefont
  {Hernandez}, \citenamefont {Loewe}, \citenamefont {Paret},\ and\
  \citenamefont {Zamora}}]{Ayala:2018wux}%
  \BibitemOpen
  \bibfield  {author} {\bibinfo {author} {\bibfnamefont {A.}~\bibnamefont
  {Ayala}}, \bibinfo {author} {\bibfnamefont {C.~A.}\ \bibnamefont
  {Dominguez}}, \bibinfo {author} {\bibfnamefont {S.}~\bibnamefont
  {Hernandez-Ortiz}}, \bibinfo {author} {\bibfnamefont {L.~A.}\ \bibnamefont
  {Hernandez}}, \bibinfo {author} {\bibfnamefont {M.}~\bibnamefont {Loewe}},
  \bibinfo {author} {\bibfnamefont {D.~M.}\ \bibnamefont {Paret}}, \ and\
  \bibinfo {author} {\bibfnamefont {R.}~\bibnamefont {Zamora}},\ }\href
  {\doibase 10.1103/PhysRevD.98.031501} {\bibfield  {journal} {\bibinfo
  {journal} {Phys. Rev. D}\ }\textbf {\bibinfo {volume} {98}},\ \bibinfo
  {pages} {031501(R)} (\bibinfo {year} {2018})},\ \Eprint
  {http://arxiv.org/abs/1805.08198} {arXiv:1805.08198 [hep-ph]} \BibitemShut
  {NoStop}%
\bibitem [{\citenamefont {Shovkovy}(2013)}]{Shovkovy:2012zn}%
  \BibitemOpen
  \bibfield  {author} {\bibinfo {author} {\bibfnamefont {I.~A.}\ \bibnamefont
  {Shovkovy}},\ }\href {\doibase 10.1007/978-3-642-37305-3_2} {\bibfield
  {journal} {\bibinfo  {journal} {Lect. Notes Phys.}\ }\textbf {\bibinfo
  {volume} {871}},\ \bibinfo {pages} {13} (\bibinfo {year} {2013})},\ \Eprint
  {http://arxiv.org/abs/1207.5081} {arXiv:1207.5081 [hep-ph]} \BibitemShut
  {NoStop}%
\bibitem [{\citenamefont {Huang}\ \emph {et~al.}(2010)\citenamefont {Huang},
  \citenamefont {Huang}, \citenamefont {Rischke},\ and\ \citenamefont
  {Sedrakian}}]{Huang2010}%
  \BibitemOpen
  \bibfield  {author} {\bibinfo {author} {\bibfnamefont {X.-G.}\ \bibnamefont
  {Huang}}, \bibinfo {author} {\bibfnamefont {M.}~\bibnamefont {Huang}},
  \bibinfo {author} {\bibfnamefont {D.~H.}\ \bibnamefont {Rischke}}, \ and\
  \bibinfo {author} {\bibfnamefont {A.}~\bibnamefont {Sedrakian}},\ }\href
  {\doibase 10.1103/physrevd.81.045015} {\bibfield  {journal} {\bibinfo
  {journal} {Physical Review D}\ }\textbf {\bibinfo {volume} {81}},\ \bibinfo
  {pages} {045015} (\bibinfo {year} {2010})}\BibitemShut {NoStop}%
\end{thebibliography}%
\end{document}